\documentclass[%
 reprint,
 amsmath,amssymb,
 aps,
]{revtex4-1}
\usepackage{graphicx}
\usepackage{dcolumn}
\usepackage{bm}

\begin{document}

\preprint{APS/123-QED}

\title{Kaon-nucleon scattering to one-loop order in heavy baryon chiral perturbation theory}
\author{Bo-Lin Huang}
\email{bolin.huang@foxmail.com}
\author{Yun-De Li}%
 \email{yndxlyd@163.com}
\affiliation{%
 Department of Physics, Yunnan University, Kunming 650091, China
}%
\date{\today}

\begin{abstract}
We calculate the T-matrices of kaon-nucleon ($KN$) and antikaon-nucleon ($\overline{K}N$) scattering to one-loop order in SU(3) heavy baryon chiral perturbation theory (HB$\chi$PT). The low-energy constants (LECs) and their combinations are then determined by fitting the phase shifts of $KN$ scattering and the corresponding data. This leads to a good description of the phase shifts below 200 MeV kaon laboratory momentum. We obtain the LEC uncertainties through statistical regression analysis. We also determine the LECs through the use of scattering lengths in order to check the consistency of the HB$\chi$PT framework for different observables and obtain a consistent result. By using these LECs, we predict the $\overline{K}N$ elastic scattering phase shifts and obtain reasonable results. The scattering lengths are also predicted, which turn out to be in good agreement with the empirical values except for the isospin-0 $\overline{K}N$ scattering length that is strongly affected by the $\Lambda(1405)$ resonance. As most calculations in the chiral perturbation theory, the convergence issue is discussed in detail. Our calculations provide a possibility to investigate the baryon-baryon interaction in HB$\chi$PT.
\begin{description}
\item[PACS numbers]
13.75.Jz,12.39.Fe,12.38.Bx
\end{description}
\end{abstract}

\pacs{Valid PACS appear here}
\maketitle


\section{Introduction}
Chiral perturbation theory (ChPT) is the effective field theory of quantum chromodynamics (QCD) at energies below the scale of chiral symmetry breaking $\Lambda_{\chi}\sim1$ GeV\cite{mach2011,sche2012}. As we all know, the relativistic framework for baryons in ChPT does not naturally provide a simple power-counting scheme as for mesons because of the baryon mass, which does not vanish in the chiral limit. Relativistic (such as infrared regularization\cite{bech1999} and the extended on-mass-shell scheme\cite{gege1999,fuch2003}) and heavy baryon\cite{jenk1991,bern1992} approaches have been proposed and developed to solve the power-counting problem. Recently, the relativistic approaches have made some progress. For some observables, the chiral series even show a better convergence than the heavy baryon approach\cite{ren2012,alar2013}. However, the heavy baryon chiral perturbation theory (HB$\chi$PT) is still a reasonable and useful tool in the study of the meson-baryon scattering. The expansion in HB$\chi$PT is expanded simultaneously in terms of $p/\Lambda_{\chi}$ and $p/M_{0}$, where $p$ represents the meson momentum or its mass or the small residue momentum of baryon in the nonrelativistic limit and $M_{0}$ denotes the baryon mass in the chiral limit.

Over the years, the low-energy processes have been widely investigated in the SU(2) HB$\chi$PT. Fettes \textit{et al}. have investigated pion-nucleon scattering up to the fourth order\cite{fett1998,fett2000}. The low-energy constants (LECs) of the SU(2) chiral pion-nucleon Lagrangian were determined by fitting various empirical phase shifts. The threshold parameters were also predicted in Refs.~\cite{fett1998,fett2000}. Krebs, Gasparyan, and Epelbaum calculated the chiral three-nucleon force at fifth order by using the LECs from $\pi N$ scattering at fourth order\cite{kreb2012}, and Entem \textit{et al}. considered peripheral nucleon-nucleon scattering at fifth order through using these LECs\cite{ente2015}. These predictions are in good agreement with the data.

For processes involving kaons or hyperons, the situation is more complicated. One has to use the SU(3) HB$\chi$PT in comparison to the SU(2) sector of $\pi N$ scattering. These involve several new problems. First, there are more unknown LECs needed to be determined through experimental data which are insufficient at present. Second, the kaon mass $m_{K}$ is larger than the pion mass $m_{\pi}$ duo to broken SU(3) symmetry. In fact, the pertinent expansion parameter $m_{K}/\Lambda_{\chi}\sim1/2$ results in a low convergence rate. Third, the $\overline{K}N$ and $KN$ scattering are inelastic and elastic at low energies, respectively. These involve inconsistent predictions duo to the dynamical differences between $KN$ and $\overline{K}N$ scattering. However, Kaiser achieved some success when analyzing the $KN$ and $\overline{K}N$ scattering lengths in SU(3) HB$\chi$PT\cite{kais2001}. Then Liu and Zhu generalized this method to the predictions of meson-baryon scattering lengths\cite{liu20071,liu20072,liu2011,liu2012}. They obtained reasonable results. But higher-order corrections are needed to consider due to the complicated convergence. That leads to involving more LECs and needs more experimental meson-baryon scattering lengths which are unavailable for now. In this paper, we will determine the LECs by fitting the phase shifts of the elastic $KN$ scattering and make predictions up to one-loop order, as the $\pi N$ scattering in the framework of SU(2) HB$\chi$PT. 

In Sec.~\ref{lagrangian}, we summarize the Lagrangians involved in the evaluation up to one-loop order contributions. In Sec.~\ref{tmatrix}, we present the T-matrices of the elastic $KN$ and $\overline{K}N$ scattering. In Sec.~\ref{phase} we explain how we calculate the phase shifts and the scattering lengths. Section~\ref{results} contains the results and discussions and also includes a brief summary. Appendix~\ref{One-loop amplitudes} contains the amplitudes from one-loop diagrams. Apppendix~\ref{Threshold T-matrices} contains the threshold T-matrices and the relation between the threshold T-matrices with the s-wave scattering lengths.

\section{Lagrangian}
\label{lagrangian}
Our calculation of the elastic $KN$ and $\overline{K}N$ scattering is based on the effective SU(3) chiral Lagrangian in HB$\chi$PT
\begin{eqnarray}
\label{eq1}
\mathcal{L}=\mathcal{L}_{\phi\phi}+\mathcal{L}_{\phi B}.
\end{eqnarray}
Here, the SU(3) matrix $\phi$ and $B$ represent the pseudoscalar Goldstone fields ($\phi=\pi,K,\overline{K},\eta$) and the octet baryons fields, respectively. The lowest-order effective SU(3) chiral Lagrangians for meson-meson and meson-baryon interaction takes the form\cite{bora1997}
\begin{eqnarray}
\label{eq2}
\mathcal{L}_{\phi\phi}^{(2)}=\frac{f^{2}}{4}\text{tr}(u_{\mu}u^{\mu}+\chi_{+}),
\end{eqnarray}
\begin{eqnarray}
\label{eq3}
 \mathcal{L}_{\phi B}^{(1)}&=&\text{tr}(i\overline{B}[v\cdot D,B])+D\text{tr}(\overline{B}S_{\mu}\{u^{\mu},B\})\nonumber\\
&&+F\text{tr}(\overline{B}S_{\mu}[u^{\mu},B]),
\end{eqnarray}
where $D_{\mu}$ denotes the covariant derivative
\begin{eqnarray}
\label{eq4}
[D_{\mu},B]=\partial_{\mu}B+[\Gamma_{\mu},B]
\end{eqnarray}
and $S_{\mu}$ is the covariant spin operator \textit{\`a la} Pauli-Lubanski
\begin{eqnarray}
\label{eq5}
S_{\mu}=\frac{i}{2}\gamma_{5}\sigma_{\mu\nu}v^{\nu},
\end{eqnarray}
with
\begin{eqnarray}
\label{eq6}
[S_{\mu},S_{\nu}]=i\epsilon_{\mu\nu\sigma\rho}v^{\sigma}S^{\rho},\,\,\{S_{\mu},S_{\nu}\}=\frac{1}{2}(v_{\mu}v_{\nu}-g_{\mu\nu}),
\end{eqnarray}
where $\epsilon_{\mu\nu\sigma\rho}$ is the completely antisymmetric tensor in four indices, $\epsilon_{0123}=1$. The chiral connection $\Gamma^{\mu}=[\xi^{\dag},\partial^{\mu}\xi]/2$ and the axial vector quantity $u^{\mu}=i\{\xi^{\dag},\partial^{\mu}\xi\}$ contain even number meson fields and odd number meson fields, respectively. The SU(3) matrix $U=\xi^{2}=\text{exp}(i\phi/f)$ collects the pseudoscalar Goldstone fields. The parameter $f$ is the pseudoscalar decay constant in the chiral limit. The axial vector coupling constants $D$ and $F$ can be determined by fitting the semileptonic decays ($D=0.80,F=0.50$)\cite{bora1999}. The combination $\chi_{+}=\xi^{\dag}\chi\xi^{\dag}+\xi\chi\xi$ with $\chi=\text{diag}(m_{\pi}^{2},m_{\pi}^{2},2m_{K}^{2}-m_{\pi}^{2})$ results in explicit chiral symmetry breaking. The complete heavy baryon Lagrangian  at next-to-leading order can be written as
\begin{eqnarray}
\label{eq7}
\mathcal{L}_{\phi B}^{(2)}=\mathcal{L}_{\phi B}^{(2,1/M_{0})}+\mathcal{L}_{\phi B}^{(2,\text{ct})},
\end{eqnarray}
where $\mathcal{L}_{\phi B}^{(2,1/M_{0})}$ denotes $1/M_{0}$ corrections of dimension two with fixed coefficients and stems from the $1/M_{0}$ expansion of the original relativistic leading-order Lagrangian $\mathcal{L}_{\phi B}^{(1)}$\cite{bora1997}. These read
\begin{eqnarray}
\label{eq8}
\mathcal{L}_{\phi B}^{(2,1/M_{0})}&=&\frac{D^{2}-3F^{2}}{24M_{0}}\text{tr}(\overline{B}[v\cdot u,[v\cdot u,B]])\nonumber\\
&&-\frac{D^{2}}{12M_{0}}\text{tr}(\overline{B}B)\text{tr}(v\cdot u\,\, v\cdot u)\nonumber\\
&&-\frac{DF}{4M_{0}}\text{tr}(\overline{B}[v\cdot u,\{v\cdot u,B\}])\nonumber\\
&&-\frac{1}{2M_{0}}\text{tr}(\overline{B}[D_{\mu},[D^{\mu},B]])\nonumber\\
&&+\frac{1}{2M_{0}}\text{tr}(\overline{B}[v\cdot D,[v\cdot D,B]])\nonumber\\
&&-\frac{iD}{2M_{0}}\text{tr}(\overline{B}S_{\mu}[D^{\mu},\{v\cdot u,B\}])\nonumber\\
&&-\frac{iF}{2M_{0}}\text{tr}(\overline{B}S_{\mu}[D^{\mu},[v\cdot u,B]])\nonumber\\
&&-\frac{iF}{2M_{0}}\text{tr}(\overline{B}S_{\mu}[v\cdot u,[D^{\mu},B]])\nonumber\\
&&-\frac{iD}{2M_{0}}\text{tr}(\overline{B}S_{\mu}\{v\cdot u,[D^{\mu},B]\}),
\end{eqnarray}
where $M_{0}$ denotes the baryon mass in the chiral limit. The remaining heavy baryon Lagrangian $\mathcal{L}_{\phi B}^{(2,\text{ct})}$ proportional to the low-energy constants can be obtained from the relativistic effective meson-baryon chiral Lagrangian\cite{olle2006}
\begin{eqnarray}
\label{eq9}
\mathcal{L}_{\phi B}^{(2,\text{ct})}&=&b_{D}\text{tr}(\overline{B}\{\chi_{+},B\})+b_{F}\text{tr}(\overline{B}[\chi_{+},B])\nonumber\\
&&+b_{0}\text{tr}(\overline{B}B)\text{tr}(\chi_{+})+b_{1}\text{tr}(\overline{B}\{u^{\mu}u_{\mu},B\})\nonumber\\
&&+b_{2}\text{tr}(\overline{B}[u^{\mu}u_{\mu},B])+b_{3}\text{tr}(\overline{B}B)\text{tr}(u^{\mu}u_{\mu})\nonumber\\
&&+b_{4}\text{tr}(\overline{B}u^{\mu})\text{tr}(Bu_{\mu})+b_{5}\text{tr}(\overline{B}\{v\cdot u\,\,v\cdot u,B\})\nonumber\\
&&+b_{6}\text{tr}(\overline{B}[v\cdot u\,\,v\cdot u,B])+b_{7}\text{tr}(\overline{B}B)\text{tr}(v\cdot u\,\,v\cdot u)\nonumber\\
&&+b_{8}\text{tr}(\overline{B}v\cdot u)\text{tr}(B v\cdot u)\nonumber\\
&&+b_{9}\text{tr}(\overline{B}\{[u^{\mu},u^{\nu}],[S_{\mu},S_{\nu}]B\})\nonumber\\
&&+b_{10}\text{tr}(\overline{B}[[u^{\mu},u^{\nu}],[S_{\mu},S_{\nu}]B])\nonumber\\
&&+b_{11}\text{tr}(\overline{B}u^{\mu})\text{tr}(u^{\nu}[S_{\mu},S_{\nu}]B).
\end{eqnarray}
The first three terms proportional to the LECs $b_{D,F,0}$ result in explicit symmetry breaking. Notice that the LECs $b_{i} (i=D,F,0,1,2,3,4,5,6,7,8,9,10,11)$ have dimension $\text{mass}^{-1}$.

\section{T-Matrices}
\label{tmatrix}
We are considering only elastic kaon-nucleon and antikaon-nucleon scattering $\{K(\bm{q}),\overline{K}(\bm{q})\}+N(\bm{-q}) \rightarrow\{K(\bm{q}'),\overline{K}(\bm{q}')\}+N(\bm{-q}')$ in the center-of-momentum system (CMS) with $q=|\bm{q}|=|\bm{q}'|$. The T-matrix takes the following form:
\begin{eqnarray}
\label{eq10}
 T_{KN,\overline{K}N}^{(I)}&=&(\frac{E_{N}+M_{N}}{2M_{N}})\{V_{KN,\overline{K}N}^{(I)}(q)\nonumber\\
&&+i\bm{\sigma}\cdot(\bm{q}'\times\bm{q})W_{KN,\overline{K}N}^{(I)}(q)\},
\end{eqnarray}
with $M_{N}$ the nucleon mass, $E_{N}=(q^{2}+M_{N}^{2})^{1/2}$ the nucleon energy, and $I$ the total isospin of the kaon-nucleon system. Furthermore, $V_{KN,\overline{K}N}^{(I)}(q)$ refers to the non-spin-flip kaon-nucleon or antikaon-nucleon amplitude, and $W_{KN,\overline{K}N}^{(I)}(q)$ refers to the spin-flip kaon-nucleon or antikaon-nucleon amplitude.

\begin{figure}[t]
\includegraphics[height=5cm,width=8cm]{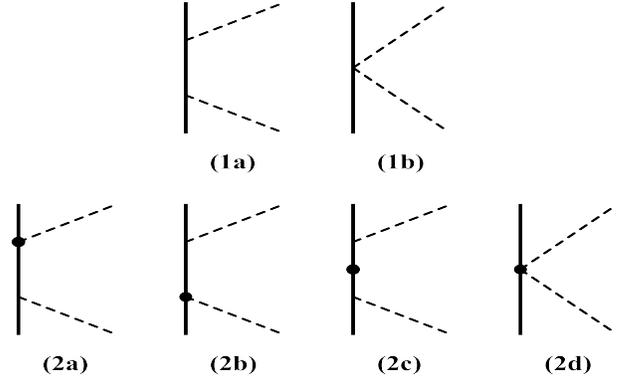}
\caption{\label{fig:treefeynman}Tree diagrams contributing to the first and second chiral orders. Dashed lines represent Goldstone bosons, and solid lines represent octet baryons. The heavy dots refer to insertions from $\mathcal{L}_{\phi B}^{(2)}$. Crossed diagrams are not shown. } 
\end{figure}

Now, we calculate the T-matrices order by order. Note that we choose $v^{\mu}=(1,0,0,0)$ for the sake of convenience throughout this paper. The leading-order $\mathcal{O}(q)$ amplitudes corresponding to diagrams (1a) and (1b) in Fig.~\ref{fig:treefeynman} (including also the crossed diagram) read
\begin{eqnarray}
\label{eq11}
V_{KN}^{(1)}(q)=\frac{1}{3f_{K}^{2}}[-3w+(D^{2}+3F^{2})\frac{q^{2}z}{w}],
\end{eqnarray}
\begin{eqnarray}
\label{eq12}
W_{KN}^{(1)}(q)=-\frac{D^{2}+3F^{2}}{3wf_{K}^{2}},
\end{eqnarray}
\begin{eqnarray}
\label{eq13}
V_{KN}^{(0)}(q)=\frac{(2D^{2}-6DF)q^{2}z}{3wf_{K}^{2}},
\end{eqnarray}
\begin{eqnarray}
\label{eq14}
W_{KN}^{(0)}(q)=-\frac{2D^{2}-6DF}{3wf_{K}^{2}},
\end{eqnarray}
\begin{eqnarray}
\label{eq15}
V_{\overline{K}N}^{(1)}(q)=\frac{1}{2f_{K}^{2}}[w-(D-F)^{2}\frac{q^{2}z}{w}],
\end{eqnarray}
\begin{eqnarray}
\label{eq16}
W_{\overline{K}N}^{(1)}(q)=-\frac{(D-F)^{2}}{2wf_{K}^{2}},
\end{eqnarray}
\begin{eqnarray}
\label{eq17}
V_{\overline{K}N}^{(0)}(q)=\frac{1}{6f_{K}^{2}}[9w-(D+3F)^{2}\frac{q^{2}z}{w}],
\end{eqnarray}
\begin{eqnarray}
\label{eq18}
W_{\overline{K}N}^{(0)}(q)=-\frac{(D+3F)^{2}}{6wf_{K}^{2}},
\end{eqnarray}
where $w=(m_{K}^{2}+q^{2})^{1/2}$ denotes the kaon CMS energy and $z=\text{cos}(\theta)$ the angular variable between $\bm{q}$ and $\bm{q}'$. We also take the renormalized kaon decay constant $f_{K}$ instead of $f$ (the chiral limit value).

At next-to-leading order $\mathcal{O}(q^{2})$, one has the contribution from the second row diagrams of Fig.~\ref{fig:treefeynman} (including also the crossed diagrams) involving the vertices from the $\mathcal{O}(q^{2})$ Lagrangian $\mathcal{L}_{\phi B}^{(2,1/M_{0})}$ and $\mathcal{L}_{\phi B}^{(2,\text{ct})}$. First, for the vertices from the $\mathcal{L}_{\phi B}^{(2,1/M_{0})}$, we have
\begin{eqnarray}
\label{eq19}
V_{KN}^{(1)}(q)&=&\frac{1}{6M_{0}f_{K}^{2}}(D^{2}+3F^{2})[-w^{2}+2(z+2)q^{2}\nonumber\\&&-\frac{3(1+z)}{D^{2}+3F^{2}}q^{2}-2z(1+z)\frac{q^{4}}{w^{2}}],
\end{eqnarray}
\begin{eqnarray}
\label{eq20}
W_{KN}^{(1)}(q)&=&-\frac{1}{3M_{0}f_{K}^{2}}(D^{2}+3F^{2})
[1-(1+z)\frac{q^{2}}{w^2}],
\end{eqnarray}
\begin{eqnarray}
\label{eq21}
V_{KN}^{(0)}&=&\frac{1}{3M_{0}f_{K}^{2}}(D^{2}-3DF)[-w^{2}+2(z+2)q^{2}\nonumber\\&&-2z(1+z)\frac{q^{4}}{w^{2}}],
\end{eqnarray}
\begin{eqnarray}
\label{eq22}
W_{KN}^{(0)}=-\frac{1}{3M_{0}f_{K}^{2}}(2D^{2}-6DF)[1-(1+z)\frac{q^{2}}{w^{2}}],
\end{eqnarray}
\begin{eqnarray}
\label{eq23}
V_{\overline{K}N}^{(1)}&=&-\frac{1}{4M_{0}f_{K}^{2}}[(D-F)^{2}w^{2}+2z(D-F)^{2}q^{2}\nonumber\\&&-(1+z)q^{2}],
\end{eqnarray}
\begin{eqnarray}
\label{eq24}
W_{\overline{K}N}^{(1)}(q)=-\frac{(D-F)^{2}}{2M_{0}f_{K}^{2}},
\end{eqnarray}

\begin{eqnarray}
\label{eq25}
V_{\overline{K}N}^{(0)}(q)&=&-\frac{1}{12M_{0}f_{K}^{2}}[(D+3F)^{2}w^{2}\nonumber\\&&+2z(D+3F)^{2}q^{2}-9(1+z)q^{2}],
\end{eqnarray}
\begin{eqnarray}
\label{eq26}
W_{\overline{K}N}^{(0)}(q)=-\frac{(D+3F)^{2}}{6M_{0}f_{K}^{2}}.
\end{eqnarray}
Second, for the vertices from the $\mathcal{L}_{\phi B}^{(2,\text{ct})}$, we introduce
\begin{eqnarray}
\label{eq27}
&&\alpha_{\eta}=4b_{D}m_{\eta}^{2}+3b_{0}(m_{\pi}^{2}+m_{\eta}^{2}),\nonumber\\
&&\alpha_{\pi}=4b_{D}m_{\pi}^{2}+3b_{0}(m_{\pi}^{2}+m_{\eta}^{2})
\end{eqnarray}
to make the following expressions more compact. The amplitudes read
\begin{eqnarray}
\label{eq28}
V_{KN}^{(1)}&=&-\frac{1}{f_{K}^{2}}[4(b_{D}+b_{0})m_{K}^{2}+(C_{1}+C_{2})w^{2}-C_{1}zq^{2}]\nonumber\\
&&+\frac{zq^{2}}{12w^{2}f_{K}^{2}}[(D+3F)^{2}\alpha_{\eta}+3(D-F)^{2}\alpha_{\pi}],
\end{eqnarray}
\begin{eqnarray}
\label{eq29}
W_{KN}^{(1)}&=&-\frac{1}{f_{K}^{2}}C_{3}-\frac{1}{12w^{2}f_{K}^{2}}[(D+3F)^{2}\alpha_{\eta}\nonumber\\&&+3(D-F)^{2}\alpha_{\pi}],
\end{eqnarray}
\begin{eqnarray}
\label{eq30}
V_{KN}^{(0)}&=&\frac{1}{f_{K}^{2}}[4(b_{F}-b_{0})m_{K}^{2}+(C_{4}+C_{5})w^{2}-C_{4}zq^{2}]\nonumber\\&&+\frac{zq^{2}}{12w^{2}f_{K}^{2}}[9(D-F)^{2}\alpha_{\pi}-(D+3F)^{2}\alpha_{\eta}],
\end{eqnarray}
\begin{eqnarray}
\label{eq31}
W_{KN}^{(0)}&=&-\frac{1}{f_{K}^{2}}C_{6}-\frac{1}{12w^{2}f_{K}^{2}}[9(D-F)^{2}\alpha_{\pi}\nonumber\\
&&-(D+3F)^{2}\alpha_{\eta}],
\end{eqnarray}
\begin{eqnarray}
\label{eq32}
V_{\overline{K}N}^{(1)}&=&\frac{1}{f_{K}^{2}}[(2b_{F}-2b_{D}-4b_{0})m_{K}^{2}-\frac{1}{2}(C_{1}+C_{2}\nonumber\\
&&-C_{4}-C_{5})w^{2}+\frac{1}{2}(C_{1}-C_{4})zq^{2}]\nonumber\\
&&+\frac{zq^{2}}{2w^{2}f_{K}^{2}}(D-F)^{2}\alpha_{\pi},
\end{eqnarray}
\begin{eqnarray}
\label{eq33}
W_{\overline{K}N}^{(1)}=\frac{1}{2f_{K}^{2}}(C_{3}+C_{6})+\frac{1}{2w^{2}f_{K}^{2}}(D-F)^{2}\alpha_{\pi},
\end{eqnarray}
\begin{eqnarray}
\label{eq34}
V_{\overline{K}N}^{(0)}&=&-\frac{1}{f_{K}^{2}}[2(3b_{D}+b_{F}+2b_{0})m_{K}^{2}+\frac{1}{2}(3C_{1}+3C_{2}\nonumber\\
&&+C_{4}+C_{5})w^{2}-\frac{1}{2}(3C_{1}+C_{4})zq^{2}]\nonumber\\
&&+\frac{zq^{2}}{6w^{2}f_{K}^{2}}(D+3F)^{2}\alpha_{\eta},
\end{eqnarray}
\begin{eqnarray}
\label{eq35}
W_{\overline{K}N}^{(0)}=\frac{1}{2f_{K}^{2}}(3C_{3}-C_{6})+\frac{1}{6w^{2}f_{K}^{2}}(D+3F)^{2}\alpha_{\eta},
\end{eqnarray}
where
\begin{eqnarray}
\label{eqconstants}
&&C_{1}=-4b_{1}-4b_{3}-2b_{4},\nonumber\\
&&C_{2}=-4b_{5}-4b_{7}-2b_{8},\nonumber\\
&&C_{3}=4b_{10}+b_{11},\nonumber\\
&&C_{4}=-4b_{2}+4b_{3}-2b_{4},\nonumber\\
&&C_{5}=-4b_{6}+4b_{7}-2b_{8},\nonumber\\
&&C_{6}=-4b_{9}-b_{11}.
\end{eqnarray}
The six combinations of LECs $C_{i}(i=1,2,3,4,5,6)$ are introduced in order to reduce the number of LECs.
\begin{figure}[t]
\includegraphics[height=11.7cm,width=8cm]{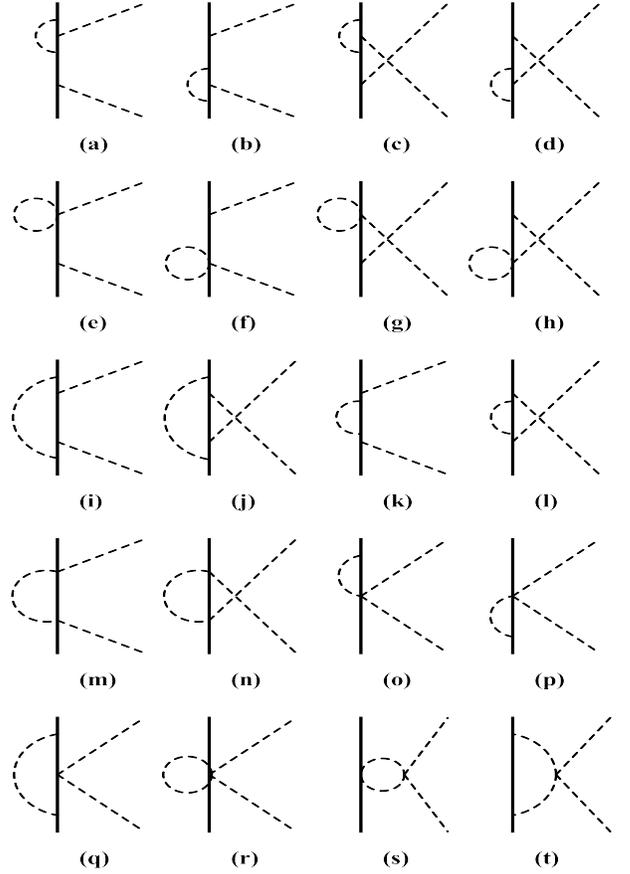}
\caption{\label{fig:knloopfeynman}Nonvanishing one-loop diagrams contributing to the third chiral order. Diagrams for external legs renormalization are not shown.  } 
\end{figure}

At the third order $\mathcal{O}(q^{3})$, we have the one-loop diagram contributions and the counterterm contributions. The nonvanishing one-loop diagrams generated by the vertices of $\mathcal{L}_{\phi\phi}^{(2)}$ and $\mathcal{L}_{\phi B}^{(1)}$ are shown in Fig.~\ref{fig:knloopfeynman}. The counterterm contribution estimated from resonance exchange was found to be much smaller than the chiral loop contribution in the case of threshold $\pi N$ scattering\cite{bern1993,bern1995}. Kaiser assumed that similar features hold for threshold $KN$ and $\bar{K}N$ scattering and also achieved some success\cite{kais2001}. Liu and Zhu also ignored the counterterm contributions when they calculated meson-baryon scattering lengths\cite{liu20071}. Later, Liu and Zhu claimed that the counterterm contributions are larger than the one-loop diagrams contributions in some T-matrices in Ref. \cite{liu20072}. But, Liu and Zhu did not consider the $\Lambda(1405)$ resonance contribution when determining the LECs and their combinations in Ref. \cite{liu20072}. However, we are not considering the counterterm contributions when calculating T-matrices at $\mathcal{O}(q^{3})$ in this paper. The nonvanishing one-loop amplitudes corresponding to loop diagrams are too tedious; thus, we present these amplitudes separately in Appendix~\ref{One-loop amplitudes}. In loop calculations, we use dimensional regularization and the minimal subtraction scheme to evaluate divergent loop integrals\cite{hoof1979,bern19951,mojz1998,bouz2000,bouz2002}. We use $f_{K}$ in all loops instead of corresponding decay constants in respective loops. The difference appears at higher order. 

\section{Calculating phase shifts and Scattering lengths}
\label{phase}
The partial wave amplitudes $f_{l\pm s}^{(I)}(q)$, where $l$ refers to the orbital angular momentum and $s$ to the spin, are given in terms of the invariant amplitudes via
\begin{eqnarray}
f_{l\pm s}^{(I)}(q)&=&\frac{E_{N}+M_{N}}{16\pi(w+E_{N})}\int_{-1}^{+1}dz[V_{KN,\overline{K}N}^{(I)}(q)P_{l}(z)\nonumber\\
&&+q^{2}W_{KN,\overline{K}N}^{(I)}(q)(P_{l\pm 1}(z)-zP_{l}(z))],
\end{eqnarray}
where $P_{l}(z)$ are conventional Legendre polynomials. For the energy range considered in this paper, the phase shifts $\delta_{l\pm s}^{(I)}(q)$ are evaluated from (for discussions about the phase shifts, see Refs.~\cite{gass1991,fett1998})
\begin{eqnarray}
\delta_{l\pm s}^{(I)}(q)=\text{arctan}(q\,\text{Re}\,f_{l\pm s}^{(I)}(q)).
\end{eqnarray}
Based upon relativistic kinematics, there is a relation  between the CMS on-shell momentum $q$ and the momentum of the incident kaon in the laboratory system $q_{K}$, 
\begin{eqnarray}
&&q^{2}=\frac{M_{N}^{2}q_{K}^{2}}{m_{K}^{2}+M_{N}^{2}+2M_{N}\sqrt{m_{K}^{2}+q_{K}^{2}}}.
\end{eqnarray}

Near threshold the scattering length for s waves and the scattering volume for p waves is given by\cite{eric1988}
\begin{eqnarray}
\label{scattering length}
a_{l\pm s}^{(I)}=\lim\limits_{q \rightarrow 0}q^{-2l-1}\,\text{tan}\,\delta_{l\pm s}^{(I)}(q).
\end{eqnarray}

\section{Results and Discussion}
\label{results}
Before calculating the phase shifts and the threshold parameters, we have to determine the LECs. There are 14 unknown LECs in $\mathcal{L}_{\phi B}^{(2,ct)}$ and $M_{0}$ also need to be determined. Fortunately, after the regrouping, we determine only $b_{D}, b_{F}, b_{0}, M_{0}$ and the six LEC combinations $C_{1,2,3,4,5,6}$ which were defined by Eq.~(\ref{eqconstants}). Throughout this paper, we use $m_{\pi}=139.57$ MeV, $m_{K}=493.68$ MeV, $m_{\eta}=547.86$ MeV, $M_{N}=938.9$ MeV, $f_{\pi}=92.21$ MeV, $f_{K}=111$ MeV, and $f_{\eta}=1.2f_{\pi}$\cite{pdg2014}, and for the axial vector coupling constants we use $D=0.8$ and $F=0.5$. We also take $\lambda=4\pi f_{\pi}$ as the chiral symmetry breaking scale.

We first determine $M_{0}$, $b_{D}$, $b_{F}$ and $b_{0}$ through the formulas of the octet-baryon masses and $\sigma_{\pi N}$ given in Ref. ~\cite{bern19951}. We take $f=f_{\pi,K,\eta}$ in the $\pi,K,\eta$ loops in these formulas, respectively. The baryon masses $M_{N}=938.9\pm 1.3$ MeV, $M_{\Sigma}=1193.4\pm 8.1$ MeV, $M_{\Xi}=1318.3\pm 6.9$ MeV, and $M_{\Lambda}=1115.7\pm 5.4$ MeV and the pion-nucleon ($\pi N$) $\sigma$ term $\sigma_{\pi N}=59.1\pm 3.5$\cite{hofe2015} are used to fit these four parameters. We obtain
\begin{eqnarray}
&&M_{0}=646.30\pm 47.72\,\, \text{MeV},\nonumber\\
&&b_{D}=0.043\pm 0.008\,\,\text{GeV}^{-1},\nonumber\\
&&b_{F}=-0.498\pm 0.003\,\,\text{GeV}^{-1},\nonumber\\
&&b_{0}=-1.003\pm 0.047\,\,\text{GeV}^{-1}
\end{eqnarray}
with $\chi^{2}/\text{d.o.f.}\sim 1.08$. In our fitting, the new $\sigma_{\pi N}$ from Ref.~\cite{hofe2015} is taken; thus, we obtain different values than those in Ref.~\cite{liu20071}. Note that the uncertainty of the $i$th LEC (here, refers to one of the $M_{0}$, $b_{D}$, $b_{F}$ and $b_{0}$) is purely the statistical uncertainty that is a measure of how much this particular parameter can change while maintaining a good description of the fitted data, as detailed in Refs.~\cite{doba2014,carl2015}.

\begin{figure}[t]
\includegraphics[height=11.25cm,width=8.5cm]{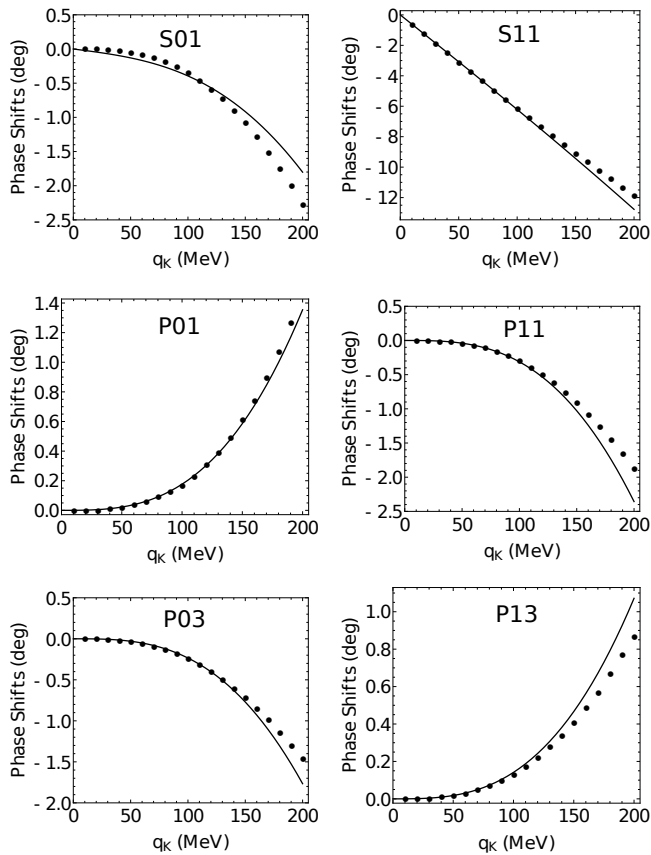}
\caption{\label{fig:KNSPWAVA}Fits and predictions for the SP92 phase shifts versus the kaon laboratory momentum $q_{K}$ in $KN$ scattering. The fits in the S01, P01 and P03 waves are the data between 100 and 140 MeV (solid dots), while the S11, P11 and P13 waves between 50 and 90 MeV. For higher and lower energies, the phase shifts are predicted.}
\end{figure}

We now determine the six LEC combinations $C_{1,2,3,4,5,6}$ by using the phase shifts of the SP92 solution, GW Institute for Nuclear Studies, for kaon-nucleon ($KN$) scattering analysis\cite{SAID,hysl1992}. Since the SP92 give no uncertainties for the phase shifts, we set a common uncertainty of $\pm4\%$ to all values before the fitting procedure. For the parameters $C_{1,2,3}$, we use the data of the S11, P11 and P13 waves between 50 and 90 MeV (15 data points in total) to fit. As to the $C_{4,5,6}$, we fit the data of the S01, P01 and P03 waves at $q_{K}=100,110,120,130,140$ MeV. The resulting LECs are given by
\begin{eqnarray}
\label{C123}
&&C_{1}=1.99\pm0.11\,\, \text{GeV}^{-1},\nonumber\\
&&C_{2}=-0.45\pm0.11\,\,\text{GeV}^{-1},\nonumber\\
&&C_{3}=6.36\pm0.09\,\, \text{GeV}^{-1},
\end{eqnarray}
with $\chi^{2}/\text{d.o.f.}\sim 0.91$ and
\begin{eqnarray}
\label{C456}
&&C_{4}=3.01\pm0.21\,\,\text{GeV}^{-1},\nonumber\\
&&C_{5}=-5.10\pm0.21\,\,\text{GeV}^{-1},\nonumber\\
&&C_{6}=-5.13\pm0.12\,\,\text{GeV}^{-1},
\end{eqnarray}
with $\chi^{2}/\text{d.o.f.}\sim 2.49$. For the uncertainties, see the above description. The corresponding S- and P-wave phase shifts are shown in Fig.~\ref{fig:KNSPWAVA}. For the P01 wave, the description of the phase shifts is surprisingly good even at higher and lower energies. The remaining waves are also in good agreement with the empirical phase shifts below 150 MeV and purely overestimated at large kaon momentum. However, to sum up, we obtain a good description for these six lowest partial waves in this one-loop order calculation of the $KN$ scattering up to surprisingly large kaon momenta.

In order to check the consistency of the ChPT framework for different observables, we now determine the low-energy constants by the scattering lengths. However, there are six LEC combinations $C_{1,2,3,4,5,6}$, but only four scattering lengths $a_{KN,\overline{K}N}^{(0,1)}$ can be used. At this time, we take the threshold T-matrices to calculate the scattering lengths; see Appendix~\ref{Threshold T-matrices}. For comparison, we use the two scattering lengths $a_{KN}^{(1)}=-0.33$ and $a_{KN}^{(0)}=0.00$ from the SP92\cite{hysl1992} to determine the two LEC combinations $C_{12}=C_{1}+C_{2}$ and $C_{45}=C_{4}+C_{5}$. The resulting LECs are given by
\begin{eqnarray}
\label{C1245}
&&C_{12}=C_{1}+C_{2}=1.59\,\,\text{GeV}^{-1},\nonumber\\
&&C_{45}=C_{4}+C_{5}=-1.99\,\,\text{GeV}^{-1}.
\end{eqnarray}
The LEC combination $C_{1}+C_{2}$ determined by the phase shifts from Eq.~(\ref{C123}) is $1.54\,\,\text{GeV}^{-1}$, while the $C_{4}+C_{5}$ from Eq.~(\ref{C456}) is $-2.09\,\,\text{GeV}^{-1}$. The results are consistent with the LEC combinations determined by the scattering lengths from Eq.~(\ref{C1245}) within the limit of error.

\begin{figure}[b]
\includegraphics[height=11.25cm,width=8.5cm]{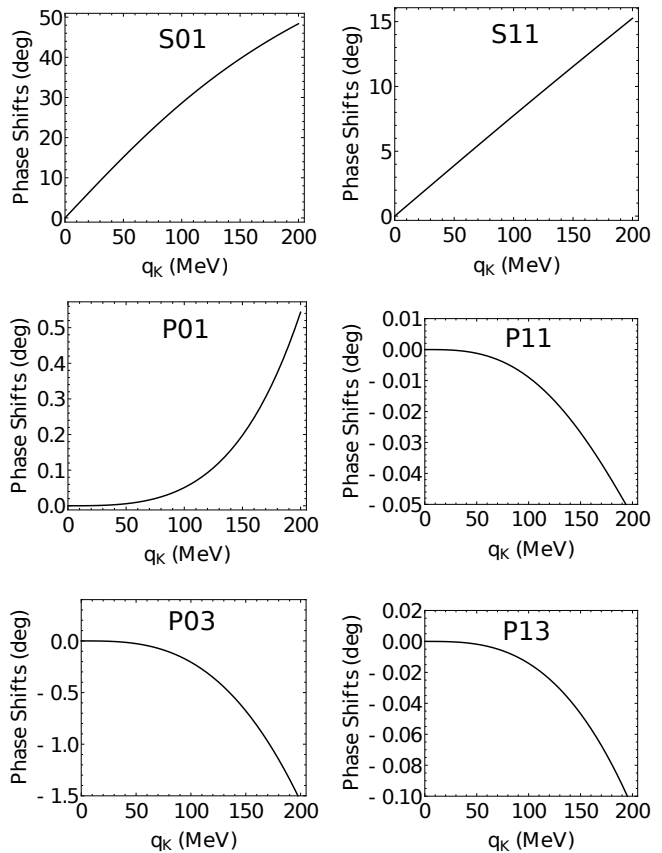}
\caption{\label{fig:KbarN_S_P_WAVA}The antikaon-nucleon ($\overline{K}N$) phase shifts versus the kaon laboratory momentum $q_{K}$. The solid lines denote our predictions through the LECs determined by the $KN$ phase shifts and the corresponding data.}
\end{figure}

In the following, we make predictions for the $\overline{K}N$ scattering through the above LECs determined by the $KN$ phase shifts and the corresponding data. At present, the existing empirical phase shifts of the $\overline{K}N$ scattering\cite{arme1969,hemi1975,gopa1977,ks20131,ks20132} are all above the kaon laboratory momentum of 200 MeV (corresponding to the CM energy of around 1460 MeV); thus, the resulting S- and P-wave phase shifts are shown in Fig.~\ref{fig:KbarN_S_P_WAVA} without the empirical phase shifts. From the plot of Fig.~\ref{fig:KbarN_S_P_WAVA}, none of the phase shifts shows the resonant behavior. The recent multichannel partial-wave analysis for $\overline{K}N$ scattering KS2013\cite{ks20131,ks20132} includes a variety of resonances, such as the S01, S11, P01, P03, P11, and P13 wave including the $\Lambda{(1670)}$, $\Sigma{(1620)}$, $\Lambda{(1710)}$, $\Lambda{(1890)}$, $\Sigma{(1770)}$ and $\Sigma{(1730)}$ resonances, respectively. But all the resonances considered by the KS2013 do not contribute to the phase shifts below the CM energy of 1460 MeV , because they are so far away. Thus, the predictions for the phase shifts of the partial waves in the $\overline{K}N$ scattering are reasonable. However, as we all know, there exists the $\Lambda{(1405)}$ resonance as a quasibound $\overline{K}N$ state below the threshold energy in a S01 wave. To solve this problem, the solution is given by the nonperturbative resummation approach with  a phenomenologically successful description of the scattering amplitude\cite{kais1995,oset1998,olle2001,lutz2002} (for a review on this issue, see Ref.~\cite{hyod2012}).

\begin{table*}
\caption{\label{tab:scattering lengths}
Values of the scattering lengths for our predictions in comparison to the various empirical values. The scattering lengths are in units of fm.}
\begin{ruledtabular}
\begin{tabular}{cccccccccccc}
Sca. Len. & Pre. A & Pre. B & SP92\footnotemark[1] & Martin\footnotemark[2] & KamanoA\footnotemark[3] & KamanoB\footnotemark[3] & TW\footnotemark[4] & TWB\footnotemark[4] & NLO\footnotemark[4]\\
\hline
$a_{KN}^{(1)}$& -0.32 &  & -0.33 &-0.33 & & \\
$a_{KN}^{(0)}$& -0.01 &  & 0.00 & 0.02 & &  \\
$a_{\overline{K}N}^{(1)}$& 0.41 & 0.41+i0.39 &  & 0.37+i0.60 & 0.07+i0.81 & 0.33+i0.49 & 0.29+i0.76 & 0.27+i0.74 & 0.57+i0.73 \\
$a_{\overline{K}N}^{(0)}$& 1.63 & 1.61+i0.23 &  & -1.70+i0.68 & -1.37+i0.67 & -1.62+i1.02 & -2.15+i0.88 & -2.15+i0.96 & -1.97+i1.05 \\
\end{tabular}
\end{ruledtabular}
\footnotetext[1]{Purely obtaining $KN$ scattering lengths, from Ref.~\cite{hysl1992}.}
\footnotetext[2]{From Ref.~\cite{mart1981}.}
\footnotetext[3]{From models A and B (denoted KamanoA and KamanoB) of Ref.~\cite{kama2014}}
\footnotetext[4]{From the three different schemes of Ref.~\cite{iked2012}, the determination of the scattering lengths included the recent precise measurement of the kaonic hydrogen by the SIDDHARTA Collaboration\cite{bazz2011}.}
\end{table*}

Now let us apply the above LECs to estimate the kaon-nucleon and antikaon-nucleon scattering lengths. We have two approaches to predict the scattering lengths. One is through the use of the Eq.~(\ref{scattering length}) and the LECs from Eqs.~(\ref{C123}) and (\ref{C456}). As before, we do not fit data below $q_{K}=50, 100$ MeV (for $C_{1,2,3}$, $C_{4,5,6}$); hence, the scattering lengths are predictions. The scattering lengths are obtained by using an incident kaon momentum $q_{K}=10$ MeV and approximating its value at the threshold. As a result, no errors  are provided. We present the values of the scattering lengths as ``Prediction A'' in Table~\ref{tab:scattering lengths}. The other is through the use of the formalism in Appendix~\ref{Threshold T-matrices} and the LECs from Eq.~(\ref{C1245}). We show the values of the scattering lengths as ``Prediction B" in Table~\ref{tab:scattering lengths}. The values purely have slightly difference than Ref.~\cite{liu20071} because different data are taken. In addition, for comparison, the various empirical values are also shown in Table~\ref{tab:scattering lengths}. We successfully predict the isospin-1 scattering lengths. For the isospin-0 $KN$ scattering length, we obtain very small negative values differing from the empirical values. However, the error will cover the difference. As expected, we fail to predict the isospin-0 $\bar{K}N$ scattering length that is dominated by the $\Lambda(1405)$ resonance. The situation is the same as the prediction for the phase shifts of the $\bar{K}N$ scattering. From this, it would be more reliable to predict the $\pi\Sigma$ \textit{et al}. scattering, although no empirical data are available. These works will be presented in our next publication.

\begin{figure}[b]
\includegraphics[height=11.25cm,width=8.5cm]{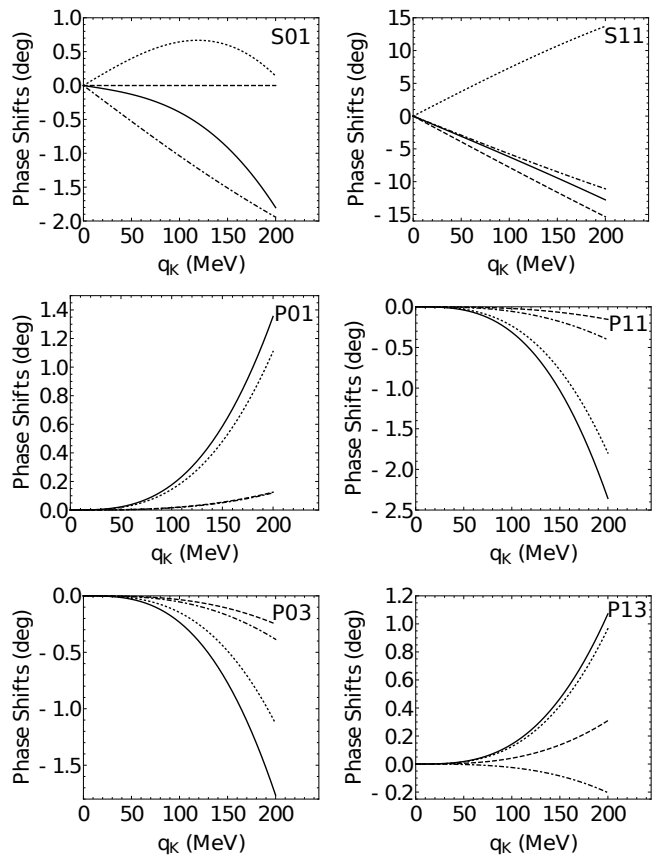}
\caption{\label{fig:KN_S_P_LO_NLO_LOOP_SUM_WAVA}Convergence properties as exemplified by our predictions for the $KN$ phase shifts. The dashed, dotted, and dashed-dotted lines denote the first, second, and third order, respectively. The solid lines give the sum of the first-, second-, and third- order contributions.}
\end{figure}

Finally, we discuss the convergence. This issue is addressed for $KN$ scattering in Fig.~\ref{fig:KN_S_P_LO_NLO_LOOP_SUM_WAVA}. For S01, the leading order is zero. The second-order contribution is much bigger than the third order and describes well the partial wave. For S11, we find that there is sizeable cancellation between the second and the third orders. This feature has also occurred in the chiral expansion of a threshold T-matrix\cite{kais2001}. For P-waves, the second order is much more important than the others in all partial waves and nearly describes well the empirical phase shifts. The situation is simpler than the $\pi N$ scattering\cite{fett1998}. According to these results, a higher-order $\mathcal{O}{(q^{4})}$ calculation is needed.

In summary, we have calculated the T-matrices for $KN$ and $\overline{K}N$ scattering to one-loop order in SU(3) HB$\chi$PT. We then fit the $\sigma_{\pi N}$, the SP92 phase shifts of $KN$ scattering, and the corresponding data to determine the LECs. This leads to a good description of the phase shifts below 200 MeV kaon momentum in the laboratory frame. We also discuss the LEC uncertainties through statistical regression analysis. In order to check the consistency of the ChPT framework for different observables, we determine the LECs by the scattering lengths, make a comparison with the LECs determined by the phase shifts, and obtain a consistent result. By using these LECs, we predict the $\overline{K}N$ scattering phase shifts and obtain a reasonable result. The s-wave scattering lengths are predicted with the energy-dependent solution (Prediction A) and in the case of the threshold T matrices (Prediction B). As expected, we fail to predict the isospin-0 $\overline{K}N$ scattering length which is dominated by the $\Lambda(1405)$ resonance. This issue can be successfully solved by the nonperturbative resummation approach, and that is not the focus of this paper. Finally, we check the convergence of the $KN$ scattering and find that the large cancellations occurred between the second and third orders in the S11 wave. In order to determine accurately the LECs and make better predictions, higher-order $\mathcal{O}(q^{4})$ calculations are needed in SU(3) HB$\chi$PT. In addition, the prediction for the octet meson and octet baryon interaction ( such as $\pi\Sigma$ and $\pi\Lambda$ scattering) will be calculated in the next publication. We also expect our calculations to provide a possibility to investigate the baryon-baryon interaction in HB$\chi$PT. 

\begin{acknowledgments}
This work is supported by the National Natural Science Foundation of China under Grants No. 11465021 and No. 11065010. B. L. H. thanks Norbert Kaiser (Technische Universit\"{a}t M\"{u}nchen), Yan-Rui Liu (Shandong University) and Jia-Qing Zhu (Yunnan University) for very helpful discussions.
\end{acknowledgments}
\appendix
\section{One-loop amplitudes}
\label{One-loop amplitudes}
In this Appendix, we present the nonvanishing amplitudes from nonvanishing one-loop diagrams. The amplitudes are shown one diagram by one diagram (but similar diagrams are grouped together) due to the expressions being too tedious. For giving the expressions as many details as possible, we use several functions in the following expressions. The normal unit step function
\begin{eqnarray}
\label{A1}
\theta(x)=
\begin{cases}
1& x>0,\\
0& x<0
\end{cases}
\end{eqnarray}
is used. We also define
\begin{eqnarray}
Q^{2}=2q^{2}(z-1),
\end{eqnarray}
\begin{eqnarray}
r(m)=\sqrt{|1-\frac{4m^{2}}{Q^{2}}|}.
\end{eqnarray}
\begin{widetext}
Figures 2(a)-2(d):
\begin{eqnarray}
\label{A2}
V_{KN}^{(1)}&=&\frac{zq^{2}}{144\pi^{2}w^{2}f_{K}^{4}}\{\alpha_{\pi}^{DF}[w^{3}-wm_{\pi}^{2}+\pi m_{\pi}^{3}+(3wm_{\pi}^{2}
-2w^{3})\text{ln}\frac{m_{\pi}}{\lambda}-2(w^{2}-m_{\pi}^{2})^{3/2}\text{ln}\frac{w+\sqrt{w^{2}-m_{\pi}^{2}}}{m_{\pi}}]\nonumber\\
&&+\alpha_{K}^{DF}[w^{3}-wm_{K}^{2}+\pi m_{K}^{3}+(3wm_{K}^{2}-2w^{3})\text{ln}\frac{m_{K}}{\lambda}
-2(w^{2}-m_{K}^{2})^{3/2}\text{ln}\frac{w+\sqrt{w^{2}-m_{K}^{2}}}{m_{K}}]\nonumber\\
&&+\alpha_{\eta}^{DF}[w^{3}-wm_{\eta}^{2}+\pi m_{\eta}^{3}+(3wm_{\eta}^{2}-2w^{3})\text{ln}\frac{m_{\eta}}{\lambda}
-2(m_{\eta}^{2}-w^{2})^{3/2}(\text{arccos}\frac{w}{m_{\eta}})\theta(m_{\eta}^{2}-w^{2} )\nonumber\\
&&-2(w^{2}-m_{\eta}^{2})^{3/2}(\text{ln}\frac{w+\sqrt{w^{2}-m_{\eta}^{2}}}{m_{\eta}})\theta(w^{2}-m_{\eta}^{2})]\},
\end{eqnarray}

\begin{eqnarray}
\label{A3}
W_{KN}^{(1)}=-\frac{V_{KN}^{(1)}}{zq^{2}},
\end{eqnarray}

\begin{eqnarray}
\label{A2}
V_{KN}^{(0)}&=&\frac{zq^{2}}{144\pi^{2}w^{2}f_{K}^{4}}\{\beta_{\pi}^{DF}[w^{3}-wm_{\pi}^{2}+\pi m_{\pi}^{3}+(3wm_{\pi}^{2}
-2w^{3})\text{ln}\frac{m_{\pi}}{\lambda}-2(w^{2}-m_{\pi}^{2})^{3/2}\text{ln}\frac{w+\sqrt{w^{2}-m_{\pi}^{2}}}{m_{\pi}}]\nonumber\\
&&+\beta_{K}^{DF}[w^{3}-wm_{K}^{2}+\pi m_{K}^{3}+(3wm_{K}^{2}-2w^{3})\text{ln}\frac{m_{K}}{\lambda}
-2(w^{2}-m_{K}^{2})^{3/2}\text{ln}\frac{w+\sqrt{w^{2}-m_{K}^{2}}}{m_{K}}]\nonumber\\
&&+\beta_{\eta}^{DF}[w^{3}-wm_{\eta}^{2}+\pi m_{\eta}^{3}+(3wm_{\eta}^{2}-2w^{3})\text{ln}\frac{m_{\eta}}{\lambda}
-2(m_{\eta}^{2}-w^{2})^{3/2}(\text{arccos}\frac{w}{m_{\eta}})\theta(m_{\eta}^{2}-w^{2} )\nonumber\\
&&-2(w^{2}-m_{\eta}^{2})^{3/2}(\text{ln}\frac{w+\sqrt{w^{2}-m_{\eta}^{2}}}{m_{\eta}})\theta(w^{2}-m_{\eta}^{2})]\},
\end{eqnarray}

\begin{eqnarray}
\label{A3}
W_{KN}^{(0)}=-\frac{V_{KN}^{(0)}}{zq^{2}},
\end{eqnarray}

\begin{eqnarray}
V_{\overline{K}N}^{(1)}&=&\frac{zq^{2}}{96\pi^{2}w^{2}f_{K}^{4}}\{\gamma_{\pi}^{DF}[wm_{\pi}^{2}-w^{3}+\pi m_{\pi}^{3}+(2w^{3}-3wm_{\pi}^{2})\text{ln}\frac{m_{\pi}}{\lambda}-2(w^{2}-m_{\pi}^{2})^{3/2}(i\pi-\text{ln}\frac{w+\sqrt{w^{2}-m_{\pi}^{2}}}{m_{\pi}})]\nonumber\\
&&+\gamma_{K}^{DF}[wm_{K}^{2}-w^{3}+\pi m_{K}^{3}+(2w^{3}-3wm_{K}^{2})\text{ln}\frac{m_{K}}{\lambda}-2(w^{2}-m_{K}^{2})^{3/2}(i\pi-\text{ln}\frac{w+\sqrt{w^{2}-m_{K}^{2}}}{m_{K}})]\nonumber\\
&&+\gamma_{\eta}^{DF}[wm_{\eta}^{2}-w^{3}+\pi m_{\eta}^{3}+(2w^{3}-3wm_{\eta}^{2})\text{ln}\frac{m_{\eta}}{\lambda}-2(m_{\eta}^{2}-w^{2})^{3/2}(\text{arccos}\frac{-w}{m_{\eta}})\theta(m_{\eta}^{2}-w^{2})\nonumber\\
&&-2(w^{2}-m_{\pi}^{2})^{3/2}(i\pi-\text{ln}\frac{w+\sqrt{w^{2}-m_{\pi}^{2}}}{m_{\pi}})\theta(w^{2}-m_{\eta}^{2})]\},
\end{eqnarray}

\begin{eqnarray}
W_{\overline{K}N}^{(1)}=\frac{V_{\overline{K}N}^{(1)}}{zq^{2}},
\end{eqnarray}

\begin{eqnarray}
V_{\overline{K}N}^{(0)}&=&\frac{zq^{2}}{144\pi^{2}w^{2}f_{K}^{4}}\{\delta_{\pi}^{DF}[wm_{\pi}^{2}-w^{3}+\pi m_{\pi}^{3}+(2w^{3}-3wm_{\pi}^{2})\text{ln}\frac{m_{\pi}}{\lambda}-2(w^{2}-m_{\pi}^{2})^{3/2}(i\pi-\text{ln}\frac{w+\sqrt{w^{2}-m_{\pi}^{2}}}{m_{\pi}})]\nonumber\\
&&+\delta_{K}^{DF}[wm_{K}^{2}-w^{3}+\pi m_{K}^{3}+(2w^{3}-3wm_{K}^{2})\text{ln}\frac{m_{K}}{\lambda}-2(w^{2}-m_{K}^{2})^{3/2}(i\pi-\text{ln}\frac{w+\sqrt{w^{2}-m_{K}^{2}}}{m_{K}})]\nonumber\\
&&+\delta_{\eta}^{DF}[wm_{\eta}^{2}-w^{3}+\pi m_{\eta}^{3}+(2w^{3}-3wm_{\eta}^{2})\text{ln}\frac{m_{\eta}}{\lambda}-2(m_{\eta}^{2}-w^{2})^{3/2}(\text{arccos}\frac{-w}{m_{\eta}})\theta(m_{\eta}^{2}-w^{2})\nonumber\\
&&-2(w^{2}-m_{\pi}^{2})^{3/2}(i\pi-\text{ln}\frac{w+\sqrt{w^{2}-m_{\pi}^{2}}}{m_{\pi}})\theta(w^{2}-m_{\eta}^{2})]\},
\end{eqnarray}

\begin{eqnarray}
W_{\overline{K}N}^{(0)}=\frac{V_{\overline{K}N}^{(0)}}{zq^{2}},
\end{eqnarray}

where
\begin{eqnarray}
&&\alpha_{\pi}^{DF}=-6D^{4}-9D^{3}F-3D^{2}F^{2}+9DF^{3}+9F^{4},\nonumber\\
&&\alpha_{K}^{DF}=D^4-30D^2F^2+45F^4,\nonumber\\
&&\alpha_{\eta}^{DF}=-D^4+9D^3F-15D^2F^2-9DF^3,\nonumber\\
&&\beta_{\pi}^{DF}=27D^{3}F-27D^{2}F^{2}-27DF^{3}+27F^{4},\nonumber\\
&&\beta_{K}^{DF}=-7D^4+18D^2F^2-27F^4,\nonumber\\
&&\beta_{\eta}^{DF}=-5D^4+21D^3F-27D^2F^2+27DF^3,\nonumber\\
&&\gamma_{\pi}^{DF}=-2D^{4}+6D^{3}F-10D^{2}F^{2}-6DF^{3}+12F^{4},\nonumber\\
&&\gamma_{K}^{DF}=-2D^{4}-4D^{2}F^{2}+6F^{4},\nonumber\\
&&\gamma_{\eta}^{DF}=-2D^{4}+10D^{3}F-14D^{2}F^{2}+6DF^{3},\nonumber\\
&&\delta_{\pi}^{DF}=-9D^{4}-27D^{3}F+9D^{2}F^{2}+27DF^{3},\nonumber\\
&&\delta_{K}^{DF}=5D^4-54D^2F^2+81F^4,\nonumber\\
&&\delta_{\eta}^{DF}=D^4+3D^3F-9D^2F^2-27DF^3.
\end{eqnarray}

Figures 2(e)-2(h):
\begin{eqnarray}
V_{KN}^{(1)}&=&-\frac{zq^{2}}{96w\pi^{2}f_{K}^{4}}(D^{2}+3F^{2})(m_{\pi}^{2}\text{ln}\frac{m_{\pi}}{\lambda}
+2m_{K}^{2}\text{ln}\frac{m_{K}}{\lambda}+m_{\eta}^{2}\text{ln}\frac{m_{\eta}}{\lambda}),
\end{eqnarray}
\begin{eqnarray}
W_{KN}^{(1)}=-\frac{V_{KN}^{(1)}}{zq^{2}},
\end{eqnarray}

\begin{eqnarray}
V_{KN}^{(0)}&=&-\frac{zq^{2}}{48w\pi^{2}f_{K}^{4}}(D^{2}-3DF)(m_{\pi}^{2}\text{ln}\frac{m_{\pi}}{\lambda}
+2m_{K}^{2}\text{ln}\frac{m_{K}}{\lambda}+m_{\eta}^{2}\text{ln}\frac{m_{\eta}}{\lambda}),
\end{eqnarray}
\begin{eqnarray}
W_{KN}^{(0)}=-\frac{V_{KN}^{(0)}}{zq^{2}},
\end{eqnarray}

\begin{eqnarray}
V_{\overline{K}N}^{(1)}&=&\frac{zq^{2}}{64w\pi^{2}f_{K}^{4}}(D-F)^{2}(m_{\pi}^{2}\text{ln}\frac{m_{\pi}}{\lambda}
+2m_{K}^{2}\text{ln}\frac{m_{K}}{\lambda}+m_{\eta}^{2}\text{ln}\frac{m_{\eta}}{\lambda}),
\end{eqnarray}
\begin{eqnarray}
W_{\overline{K}N}^{(1)}=\frac{V_{\overline{K}N}^{(1)}}{zq^{2}},
\end{eqnarray}

\begin{eqnarray}
V_{\overline{K}N}^{(0)}&=&\frac{zq^{2}}{192w\pi^{2}f_{K}^{4}}(D+3F)^{2}(m_{\pi}^{2}\text{ln}\frac{m_{\pi}}{\lambda}
+2m_{K}^{2}\text{ln}\frac{m_{K}}{\lambda}+m_{\eta}^{2}\text{ln}\frac{m_{\eta}}{\lambda}),
\end{eqnarray}
\begin{eqnarray}
W_{\overline{K}N}^{(0)}=\frac{V_{\overline{K}N}^{(0)}}{zq^{2}}.
\end{eqnarray}

Figures 2(i) and 2(j):
\begin{eqnarray}
V_{KN}^{(1)}&=&\frac{zq^{2}}{288\pi^{2}w^{2}f_{K}^{4}}\{\alpha_{\pi}^{DF}[w^{3}+\pi m_{\pi}^{3}-2w(m_\pi^2+w^2\text{ln}\frac{m_\pi}{\lambda})
-2(w^{2}-m_{\pi}^{2})^{3/2}\text{ln}\frac{w+\sqrt{w^{2}-m_{\pi}^{2}}}{m_{\pi}}]\nonumber\\
&&+\alpha_{K,1}^{DF}[w^{3}+\pi m_{K}^{3}-2w(m_K^2+w^2\text{ln}\frac{m_K}{\lambda})
-2(w^{2}-m_{K}^{2})^{3/2}\text{ln}\frac{w+\sqrt{w^{2}-m_{K}^{2}}}{m_{K}}]\nonumber\\
&&+\alpha_{K,2}^{DF}[\pi m_{K}^{3}+2wm_{K}^{2}-w^{3}+2w^{3}\text{ln}\frac{m_{K}}{\lambda}
-2(w^{2}-m_{K}^{2})^{3/2}(i\pi-\text{ln}\frac{w+\sqrt{w^{2}-m_{K}^{2}}}{m_{K}})]\nonumber\\
&&+\alpha_{\eta}^{DF}[w^{3}+\pi m_{\eta}^{3}-2w(m_\eta^2+w^2\text{ln}\frac{m_\eta}{\lambda})
-2(m_{\eta}^{2}-w^{2})^{3/2}(\text{arccos}\frac{w}{m_{\eta}})\theta(m_{\eta}^{2}-w^{2} )\nonumber\\
&&-2(w^{2}-m_{\eta}^{2})^{3/2}(\text{ln}\frac{w+\sqrt{w^{2}-m_{\eta}^{2}}}{m_{\eta}})\theta(w^{2}-m_{\eta}^{2})]\},
\end{eqnarray}
\begin{eqnarray}
W_{KN}^{(1)}&=&\frac{1}{288\pi^{2}w^{2}f_{K}^{4}}\{-\alpha_{\pi}^{DF}[w^{3}+\pi m_{\pi}^{3}-2w(m_\pi^2+w^2\text{ln}\frac{m_\pi}{\lambda})
-2(w^{2}-m_{\pi}^{2})^{3/2}\text{ln}\frac{w+\sqrt{w^{2}-m_{\pi}^{2}}}{m_{\pi}}]\nonumber\\
&&-\alpha_{K,1}^{DF}[w^{3}+\pi m_{K}^{3}-2w(m_K^2+w^2\text{ln}\frac{m_K}{\lambda})
-2(w^{2}-m_{K}^{2})^{3/2}\text{ln}\frac{w+\sqrt{w^{2}-m_{K}^{2}}}{m_{K}}]\nonumber\\
&&+\alpha_{K,2}^{DF}[\pi m_{K}^{3}+2wm_{K}^{2}-w^{3}+2w^{3}\text{ln}\frac{m_{K}}{\lambda}
-2(w^{2}-m_{K}^{2})^{3/2}(i\pi-\text{ln}\frac{w+\sqrt{w^{2}-m_{K}^{2}}}{m_{K}})]\nonumber\\
&&-\alpha_{\eta}^{DF}[w^{3}+\pi m_{\eta}^{3}-2w(m_\eta^2+w^2\text{ln}\frac{m_\eta}{\lambda})
-2(m_{\eta}^{2}-w^{2})^{3/2}(\text{arccos}\frac{w}{m_{\eta}})\theta(m_{\eta}^{2}-w^{2} )\nonumber\\
&&-2(w^{2}-m_{\eta}^{2})^{3/2}(\text{ln}\frac{w+\sqrt{w^{2}-m_{\eta}^{2}}}{m_{\eta}})\theta(w^{2}-m_{\eta}^{2})]\},
\end{eqnarray}

\begin{eqnarray}
V_{KN}^{(0)}&=&\frac{zq^{2}}{288\pi^{2}w^{2}f_{K}^{4}}\{\beta_{\pi}^{DF}[w^{3}+\pi m_{\pi}^{3}-2w(m_\pi^2+w^2\text{ln}\frac{m_\pi}{\lambda})
-2(w^{2}-m_{\pi}^{2})^{3/2}\text{ln}\frac{w+\sqrt{w^{2}-m_{\pi}^{2}}}{m_{\pi}}]\nonumber\\
&&+\beta_{K,1}^{DF}[w^{3}+\pi m_{K}^{3}-2w(m_K^2+w^2\text{ln}\frac{m_K}{\lambda})
-2(w^{2}-m_{K}^{2})^{3/2}\text{ln}\frac{w+\sqrt{w^{2}-m_{K}^{2}}}{m_{K}}]\nonumber\\
&&+\beta_{K,2}^{DF}[\pi m_{K}^{3}+2wm_{K}^{2}-w^{3}+2w^{3}\text{ln}\frac{m_{K}}{\lambda}
-2(w^{2}-m_{K}^{2})^{3/2}(i\pi-\text{ln}\frac{w+\sqrt{w^{2}-m_{K}^{2}}}{m_{K}})]\nonumber\\
&&+\beta_{\eta}^{DF}[w^{3}+\pi m_{\eta}^{3}-2w(m_\eta^2+w^2\text{ln}\frac{m_\eta}{\lambda})
-2(m_{\eta}^{2}-w^{2})^{3/2}(\text{arccos}\frac{w}{m_{\eta}})\theta(m_{\eta}^{2}-w^{2} )\nonumber\\
&&-2(w^{2}-m_{\eta}^{2})^{3/2}(\text{ln}\frac{w+\sqrt{w^{2}-m_{\eta}^{2}}}{m_{\eta}})\theta(w^{2}-m_{\eta}^{2})]\},
\end{eqnarray}
\begin{eqnarray}
W_{KN}^{(0)}&=&\frac{1}{288\pi^{2}w^{2}f_{K}^{4}}\{-\beta_{\pi}^{DF}[w^{3}+\pi m_{\pi}^{3}-2w(m_\pi^2+w^2\text{ln}\frac{m_\pi}{\lambda})
-2(w^{2}-m_{\pi}^{2})^{3/2}\text{ln}\frac{w+\sqrt{w^{2}-m_{\pi}^{2}}}{m_{\pi}}]\nonumber\\
&&-\beta_{K,1}^{DF}[w^{3}+\pi m_{K}^{3}-2w(m_K^2+w^2\text{ln}\frac{m_K}{\lambda})
-2(w^{2}-m_{K}^{2})^{3/2}\text{ln}\frac{w+\sqrt{w^{2}-m_{K}^{2}}}{m_{K}}]\nonumber\\
&&+\beta_{K,2}^{DF}[\pi m_{K}^{3}+2wm_{K}^{2}-w^{3}+2w^{3}\text{ln}\frac{m_{K}}{\lambda}
-2(w^{2}-m_{K}^{2})^{3/2}(i\pi-\text{ln}\frac{w+\sqrt{w^{2}-m_{K}^{2}}}{m_{K}})]\nonumber\\
&&-\beta_{\eta}^{DF}[w^{3}+\pi m_{\eta}^{3}-2w(m_\eta^2+w^2\text{ln}\frac{m_\eta}{\lambda})
-2(m_{\eta}^{2}-w^{2})^{3/2}(\text{arccos}\frac{w}{m_{\eta}})\theta(m_{\eta}^{2}-w^{2} )\nonumber\\
&&-2(w^{2}-m_{\eta}^{2})^{3/2}(\text{ln}\frac{w+\sqrt{w^{2}-m_{\eta}^{2}}}{m_{\eta}})\theta(w^{2}-m_{\eta}^{2})]\},
\end{eqnarray}

\begin{eqnarray}
V_{\overline{K}N}^{(1)}&=&-\frac{zq^{2}}{576\pi^{2}w^{2}f_{K}^{4}}\{\gamma_{\pi}^{DF}[\pi m_{\pi}^{3}+2wm_{\pi}^{2}-w^{3}+2w^{3}\text{ln}\frac{m_{\pi}}{\lambda}
-2(w^{2}-m_{\pi}^{2})^{3/2}(i\pi-\text{ln}\frac{w+\sqrt{w^{2}-m_{\pi}^{2}}}{m_{\pi}})]\nonumber\\
&&+\gamma_{K,1}^{DF}[\pi m_{K}^{3}+2wm_{K}^{2}-w^{3}+2w^{3}\text{ln}\frac{m_{K}}{\lambda}
-2(w^{2}-m_{K}^{2})^{3/2}(i\pi-\text{ln}\frac{w+\sqrt{w^{2}-m_{K}^{2}}}{m_{K}})]\nonumber\\
&&+\gamma_{K,2}^{DF}[w^{3}+\pi m_{K}^{3}-2w(m_K^2+w^2\text{ln}\frac{m_K}{\lambda})
-2(w^{2}-m_{K}^{2})^{3/2}\text{ln}\frac{w+\sqrt{w^{2}-m_{K}^{2}}}{m_{K}}]\nonumber\\
&&+\gamma_{\eta}^{DF}[\pi m_{\eta}^{3}+2wm_{\eta}^{2}-w^{3}+2w^{3}\text{ln}\frac{m_{\eta}}{\lambda}-2(m_{\eta}^{2}-w^{2})^{3/2}(\text{arccos}\frac{-w}{m_{\eta}})\theta(m_{\eta}^{2}-w^{2})\nonumber\\
&&-2(w^{2}-m_{\eta}^{2})^{3/2}(i\pi-\text{ln}\frac{w+\sqrt{w^{2}-m_{\eta}^{2}}}{m_{\eta}})\theta(w^{2}-m_{\eta}^{2})]\},
\end{eqnarray}
\begin{eqnarray}
W_{\overline{K}N}^{(1)}&=&-\frac{1}{576\pi^{2}w^{2}f_{K}^{4}}\{\gamma_{\pi}^{DF}[\pi m_{\pi}^{3}+2wm_{\pi}^{2}-w^{3}+2w^{3}\text{ln}\frac{m_{\pi}}{\lambda}
-2(w^{2}-m_{\pi}^{2})^{3/2}(i\pi-\text{ln}\frac{w+\sqrt{w^{2}-m_{\pi}^{2}}}{m_{\pi}})]\nonumber\\
&&+\gamma_{K,1}^{DF}[\pi m_{K}^{3}+2wm_{K}^{2}-w^{3}+2w^{3}\text{ln}\frac{m_{K}}{\lambda}
-2(w^{2}-m_{K}^{2})^{3/2}(i\pi-\text{ln}\frac{w+\sqrt{w^{2}-m_{K}^{2}}}{m_{K}})]\nonumber\\
&&-\gamma_{K,2}^{DF}[w^{3}+\pi m_{K}^{3}-2w(m_K^2+w^2\text{ln}\frac{m_K}{\lambda})
-2(w^{2}-m_{K}^{2})^{3/2}\text{ln}\frac{w+\sqrt{w^{2}-m_{K}^{2}}}{m_{K}}]\nonumber\\
&&+\gamma_{\eta}^{DF}[\pi m_{\eta}^{3}+2wm_{\eta}^{2}-w^{3}+2w^{3}\text{ln}\frac{m_{\eta}}{\lambda}-2(m_{\eta}^{2}-w^{2})^{3/2}(\text{arccos}\frac{-w}{m_{\eta}})\theta(m_{\eta}^{2}-w^{2})\nonumber\\
&&-2(w^{2}-m_{\eta}^{2})^{3/2}(i\pi-\text{ln}\frac{w+\sqrt{w^{2}-m_{\eta}^{2}}}{m_{\eta}})\theta(w^{2}-m_{\eta}^{2})]\},
\end{eqnarray}

\begin{eqnarray}
V_{\overline{K}N}^{(0)}&=&-\frac{zq^{2}}{576\pi^{2}w^{2}f_{K}^{4}}\{\delta_{\pi}^{DF}[\pi m_{\pi}^{3}+2wm_{\pi}^{2}-w^{3}+2w^{3}\text{ln}\frac{m_{\pi}}{\lambda}
-2(w^{2}-m_{\pi}^{2})^{3/2}(i\pi-\text{ln}\frac{w+\sqrt{w^{2}-m_{\pi}^{2}}}{m_{\pi}})]\nonumber\\
&&+\delta_{K,1}^{DF}[\pi m_{K}^{3}+2wm_{K}^{2}-w^{3}+2w^{3}\text{ln}\frac{m_{K}}{\lambda}
-2(w^{2}-m_{K}^{2})^{3/2}(i\pi-\text{ln}\frac{w+\sqrt{w^{2}-m_{K}^{2}}}{m_{K}})]\nonumber\\
&&+\delta_{K,2}^{DF}[w^{3}+\pi m_{K}^{3}-2w(m_K^2+w^2\text{ln}\frac{m_K}{\lambda})
-2(w^{2}-m_{K}^{2})^{3/2}\text{ln}\frac{w+\sqrt{w^{2}-m_{K}^{2}}}{m_{K}}]\nonumber\\
&&+\delta_{\eta}^{DF}[\pi m_{\eta}^{3}+2wm_{\eta}^{2}-w^{3}+2w^{3}\text{ln}\frac{m_{\eta}}{\lambda}-2(m_{\eta}^{2}-w^{2})^{3/2}(\text{arccos}\frac{-w}{m_{\eta}})\theta(m_{\eta}^{2}-w^{2})\nonumber\\
&&-2(w^{2}-m_{\eta}^{2})^{3/2}(i\pi-\text{ln}\frac{w+\sqrt{w^{2}-m_{\eta}^{2}}}{m_{\eta}})\theta(w^{2}-m_{\eta}^{2})]\},
\end{eqnarray}
\begin{eqnarray}
W_{\overline{K}N}^{(0)}&=&-\frac{1}{576\pi^{2}w^{2}f_{K}^{4}}\{\delta_{\pi}^{DF}[\pi m_{\pi}^{3}+2wm_{\pi}^{2}-w^{3}+2w^{3}\text{ln}\frac{m_{\pi}}{\lambda}
-2(w^{2}-m_{\pi}^{2})^{3/2}(i\pi-\text{ln}\frac{w+\sqrt{w^{2}-m_{\pi}^{2}}}{m_{\pi}})]\nonumber\\
&&+\delta_{K,1}^{DF}[\pi m_{K}^{3}+2wm_{K}^{2}-w^{3}+2w^{3}\text{ln}\frac{m_{K}}{\lambda}
-2(w^{2}-m_{K}^{2})^{3/2}(i\pi-\text{ln}\frac{w+\sqrt{w^{2}-m_{K}^{2}}}{m_{K}})]\nonumber\\
&&-\delta_{K,2}^{DF}[w^{3}+\pi m_{K}^{3}-2w(m_K^2+w^2\text{ln}\frac{m_K}{\lambda})
-2(w^{2}-m_{K}^{2})^{3/2}\text{ln}\frac{w+\sqrt{w^{2}-m_{K}^{2}}}{m_{K}}]\nonumber\\
&&+\delta_{\eta}^{DF}[\pi m_{\eta}^{3}+2wm_{\eta}^{2}-w^{3}+2w^{3}\text{ln}\frac{m_{\eta}}{\lambda}-2(m_{\eta}^{2}-w^{2})^{3/2}(\text{arccos}\frac{-w}{m_{\eta}})\theta(m_{\eta}^{2}-w^{2})\nonumber\\
&&-2(w^{2}-m_{\eta}^{2})^{3/2}(i\pi-\text{ln}\frac{w+\sqrt{w^{2}-m_{\eta}^{2}}}{m_{\eta}})\theta(w^{2}-m_{\eta}^{2})]\},
\end{eqnarray}
where
\begin{eqnarray}
&&\alpha_{\pi}^{DF}=-12D^{4}-6D^{3}F+6D^{2}F^{2}-18DF^{3}-18F^{4},\nonumber\\
&&\alpha_{K,1}^{DF}=-13D^4+42D^2F^2-45F^4,\nonumber\\
&&\alpha_{K,2}^{DF}=-4D^{4}-24D^{2}F^{2}-36F^{4},\nonumber\\
&&\alpha_{\eta}^{DF}=-D^{4}+6D^{3}F-12D^{2}F^{2}+18DF^{3}-27F^{4},\nonumber\\
&&\beta_{\pi}^{DF}=-9D^{4}-18D^{3}F-36D^{2}F^{2}-54DF^{3}-27F^{4},\nonumber\\
&&\beta_{K,1}^{DF}=11D^{4}-54D^{2}F^{2}+27F^{4},\nonumber\\
&&\beta_{K,2}^{DF}=-16D^{4}+96D^{3}F-144D^{2}F^{2},\nonumber\\
&&\beta_{\eta}^{DF}=-2D^{4}+18D^{3}F-54D^{2}F^{2}+54DF^{3},\nonumber\\
&&\gamma_{\pi}^{DF}=21D^{4}+24D^{3}F+30D^{2}F^{2}+72DF^{3}+45F^{4},\nonumber\\
&&\gamma_{K,1}^{DF}=2D^{4}+12D^{2}F^{2}+18F^{4},\nonumber\\
&&\gamma_{K,2}^{DF}=-20D^{4}+96D^{3}F-168D^{2}F^{2}-36F^{4},\nonumber\\
&&\gamma_{\eta}^{DF}=3D^{4}-24D^{3}F+66D^{2}F^{2}-72DF^{3}+27F^{4},\nonumber\\
&&\delta_{\pi}^{DF}=27D^{4}-54D^{2}F^{2}+27F^{4},\nonumber\\
&&\delta_{K,1}^{DF}=50D^{4}-180D^{2}F^{2}+162F^{4},\nonumber\\
&&\delta_{K,2}^{DF}=4D^{4}-96D^{3}F+72D^{2}F^{2}-108F^{4},\nonumber\\
&&\delta_{\eta}^{DF}=D^{4}-18D^{2}F^{2}+81F^{4}.
\end{eqnarray}

Figures 2(k) and 2(l):
\begin{eqnarray}
V_{KN}^{(1)}&=&-\frac{zq^{2}}{1152\pi^{2}w^{2}f_{K}^{4}}\{\alpha_{\pi}^{DF}[w^{3}-wm_{\pi}^{2}+(3wm_{\pi}^{2}
-2w^{3})\text{ln}\frac{m_{\pi}}{\lambda}
-2(w^{2}-m_{\pi}^{2})^{3/2}\text{ln}\frac{w+\sqrt{w^{2}-m_{\pi}^{2}}}{m_{\pi}}]\nonumber\\
&&+\alpha_{K}^{DF}[w^{3}-wm_{K}^{2}+(3wm_{K}^{2}-2w^{3})\text{ln}\frac{m_{K}}{\lambda}
-2(w^{2}-m_{K}^{2})^{3/2}\text{ln}\frac{w+\sqrt{w^{2}-m_{K}^{2}}}{m_{K}}]\nonumber\\
&&+\alpha_{\eta}^{DF}[w^{3}-wm_{\eta}^{2}+(3wm_{\eta}^{2}-2w^{3})\text{ln}\frac{m_{\eta}}{\lambda}
-2(m_{\eta}^{2}-w^{2})^{3/2}(\text{arccos}\frac{w}{m_{\eta}})\theta(m_{\eta}^{2}-w^{2} )\nonumber\\
&&-2(w^{2}-m_{\eta}^{2})^{3/2}(\text{ln}\frac{w+\sqrt{w^{2}-m_{\eta}^{2}}}{m_{\eta}})\theta(w^{2}-m_{\eta}^{2})]\},
\end{eqnarray}
\begin{eqnarray}
W_{KN}^{(1)}=-\frac{V_{KN}^{(1)}}{zq^{2}},
\end{eqnarray}

\begin{eqnarray}
V_{KN}^{(0)}&=&-\frac{zq^{2}}{1152\pi^{2}w^{2}f_{K}^{4}}\{\beta_{\pi}^{DF}[w^{3}-wm_{\pi}^{2}+(3wm_{\pi}^{2}
-2w^{3})\text{ln}\frac{m_{\pi}}{\lambda}
-2(w^{2}-m_{\pi}^{2})^{3/2}\text{ln}\frac{w+\sqrt{w^{2}-m_{\pi}^{2}}}{m_{\pi}}]\nonumber\\
&&+\beta_{K}^{DF}[w^{3}-wm_{K}^{2}+(3wm_{K}^{2}-2w^{3})\text{ln}\frac{m_{K}}{\lambda}
-2(w^{2}-m_{K}^{2})^{3/2}\text{ln}\frac{w+\sqrt{w^{2}-m_{K}^{2}}}{m_{K}}]\nonumber\\
&&+\beta_{\eta}^{DF}[w^{3}-wm_{\eta}^{2}+(3wm_{\eta}^{2}-2w^{3})\text{ln}\frac{m_{\eta}}{\lambda}
-2(m_{\eta}^{2}-w^{2})^{3/2}(\text{arccos}\frac{w}{m_{\eta}})\theta(m_{\eta}^{2}-w^{2} )\nonumber\\
&&-2(w^{2}-m_{\eta}^{2})^{3/2}(\text{ln}\frac{w+\sqrt{w^{2}-m_{\eta}^{2}}}{m_{\eta}})\theta(w^{2}-m_{\eta}^{2})]\},
\end{eqnarray}
\begin{eqnarray}
W_{KN}^{(0)}=-\frac{V_{KN}^{(0)}}{zq^{2}},
\end{eqnarray}

\begin{eqnarray}
V_{\overline{K}N}^{(1)}&=&\frac{zq^{2}}{96\pi^{2}w^{2}f_{K}^{4}}\{\gamma_{\pi}^{DF}[w^{3}-wm_{\pi}^{2}+(3wm_{\pi}^{2}-2w^{3})\text{ln}\frac{m_{\pi}}{\lambda}+2(w^{2}-m_{\pi}^{2})^{3/2}(i\pi-\text{ln}\frac{w+\sqrt{w^{2}-m_{\pi}^{2}}}{m_{\pi}})]\nonumber\\
&&+\gamma_{K}^{DF}[w^{3}-wm_{K}^{2}+(3wm_{K}^{2}-2w^{3})\text{ln}\frac{m_{K}}{\lambda}+2(w^{2}-m_{K}^{2})^{3/2}(i\pi-\text{ln}\frac{w+\sqrt{w^{2}-m_{K}^{2}}}{m_{K}})]\nonumber\\
&&+\gamma_{\eta}^{DF}[w^{3}-wm_{\eta}^{2}+(3wm_{\eta}^{2}-2w^{3})\text{ln}\frac{m_{\eta}}{\lambda}+2(m_{\eta}^{2}-w^{2})^{3/2}(\text{arccos}\frac{-w}{m_{\eta}})\theta(m_{\eta}^{2}-w^{2})\nonumber\\
&&+2(w^{2}-m_{\eta}^{2})^{3/2}(i\pi-\text{ln}\frac{w+\sqrt{w^{2}-m_{\eta}^{2}}}{m_{\eta}})\theta(w^{2}-m_{\eta}^{2})]\},
\end{eqnarray}
\begin{eqnarray}
W_{\overline{K}N}^{(1)}=\frac{V_{\overline{K}N}^{(1)}}{zq^{2}},
\end{eqnarray}

\begin{eqnarray}
V_{\overline{K}N}^{(0)}&=&\frac{zq^{2}}{576\pi^{2}w^{2}f_{K}^{4}}\{\delta_{\pi}^{DF}[w^{3}-wm_{\pi}^{2}+(3wm_{\pi}^{2}-2w^{3})\text{ln}\frac{m_{\pi}}{\lambda}+2(w^{2}-m_{\pi}^{2})^{3/2}(i\pi-\text{ln}\frac{w+\sqrt{w^{2}-m_{\pi}^{2}}}{m_{\pi}})]\nonumber\\
&&+\delta_{K}^{DF}[w^{3}-wm_{K}^{2}+(3wm_{K}^{2}-2w^{3})\text{ln}\frac{m_{K}}{\lambda}+2(w^{2}-m_{K}^{2})^{3/2}(i\pi-\text{ln}\frac{w+\sqrt{w^{2}-m_{K}^{2}}}{m_{K}})]\nonumber\\
&&+\delta_{\eta}^{DF}[w^{3}-wm_{\eta}^{2}+(3wm_{\eta}^{2}-2w^{3})\text{ln}\frac{m_{\eta}}{\lambda}+2(m_{\eta}^{2}-w^{2})^{3/2}(\text{arccos}\frac{-w}{m_{\eta}})\theta(m_{\eta}^{2}-w^{2})\nonumber\\
&&+2(w^{2}-m_{\eta}^{2})^{3/2}(i\pi-\text{ln}\frac{w+\sqrt{w^{2}-m_{\eta}^{2}}}{m_{\eta}})\theta(w^{2}-m_{\eta}^{2})]\},
\end{eqnarray}
\begin{eqnarray}
W_{\overline{K}N}^{(0)}=\frac{V_{\overline{K}N}^{(0)}}{zq^{2}},
\end{eqnarray}

where
\begin{eqnarray}
&&\alpha_{\pi}^{DF}=24D^{4}+48D^{3}F+192D^{2}F^{2}-144DF^{3}+72F^{4},\nonumber\\
&&\alpha_{K}^{DF}=40D^4-48D^3F+144D^2F^2+144DF^3+360F^4,\nonumber\\
&&\alpha_{\eta}^{DF}=16D^{4}+48D^{2}F^{2},\nonumber\\
&&\beta_{\pi}^{DF}=24D^{4}-144D^{3}F+144D^{2}F^{2}-432DF^{3}+216F^{4},\nonumber\\
&&\beta_{K}^{DF}=104D^4-240D^3F+144D^2F^2-432DF^3-216F^4,\nonumber\\
&&\beta_{\eta}^{DF}=32D^4-96D^3F,\nonumber\\
&&\gamma_{\pi}^{DF}=2D^{4}-4D^{3}F+14D^{2}F^{2}-24DF^{3}+12F^{4},\nonumber\\
&&\gamma_{K}^{DF}=6D^{4}-12D^{3}F+12D^{2}F^{2}-12DF^{3}+6F^{4},\nonumber\\
&&\gamma_{\eta}^{DF}=2D^{4}-4D^{3}F+2D^{2}F^{2},\nonumber\\
&&\delta_{\pi}^{DF}=12D^{4}+72D^{3}F+108D^{2}F^{2},\nonumber\\
&&\delta_{K}^{DF}=4D^4+24D^3F+72D^2F^2+216DF^3+324F^4,\nonumber\\
&&\delta_{\eta}^{DF}=4D^{4}+24D^3F+36D^2F^2.
\end{eqnarray}

Figure 2(m):
\begin{eqnarray}
 V_{KN}^{(1)}&=&\frac{1}{32\pi^{2}f_{K}^{4}}[4w^{3}-(8w^{3}-3wm_{K}^{2})\text{ln}\frac{m_{K}}{\lambda}
+8w^{2}\sqrt{w^{2}-m_{K}^{2}}(i\pi-\text{ln}\frac{w+\sqrt{w^{2}-m_{K}^{2}}}{m_{K}})],
\end{eqnarray}
\begin{eqnarray}
 V_{\overline{K}N}^{(1)}&=&\frac{1}{256\pi^{2}f_{K}^{4}}[40w^{3}-5(8w^{3}-3wm_{\pi}^{2})\text{ln}\frac{m_{\pi}}{\lambda}
-2(8w^{3}-3wm_{K}^{2})\text{ln}\frac{m_{K}}{\lambda}-3(8w^{3}-3wm_{\eta}^{2})\text{ln}\frac{m_{\eta}}{\lambda}\nonumber\\
&&+40w^{2}\sqrt{w^{2}-m_{\pi}^{2}}(i\pi-\text{ln}\frac{w+\sqrt{w^{2}-m_{\pi}^{2}}}{m_{\pi}})
+16w^{2}\sqrt{w^{2}-m_{K}^{2}}(i\pi-\text{ln}\frac{w+\sqrt{w^{2}-m_{K}^{2}}}{m_{K}})\nonumber\\
&&-24w^{2}\sqrt{m_{\eta}^{2}-w^{2}}(\text{arccos}\frac{-w}{m_{\eta}})\theta(m_{\eta}^{2}-w^{2})
+24w^{2}\sqrt{w^{2}-m_{\eta}^{2}}(i\pi-\text{ln}\frac{w+\sqrt{w^{2}-m_{\eta}^{2}}}{m_{\eta}})\theta(w^{2}-m_{\eta}^{2})],
\end{eqnarray}
\begin{eqnarray}
 V_{\overline{K}N}^{(0)}&=&\frac{1}{256\pi^{2}f_{K}^{4}}[120w^{3}-3(8w^{3}-3wm_{\pi}^{2})\text{ln}\frac{m_{\pi}}{\lambda}
-18(8w^{3}-3wm_{K}^{2})\text{ln}\frac{m_{K}}{\lambda}-9(8w^{3}-3wm_{\eta}^{2})\text{ln}\frac{m_{\eta}}{\lambda}\nonumber\\
&&+24w^{2}\sqrt{w^{2}-m_{\pi}^{2}}(i\pi-\text{ln}\frac{w+\sqrt{w^{2}-m_{\pi}^{2}}}{m_{\pi}})
+144w^{2}\sqrt{w^{2}-m_{K}^{2}}(i\pi-\text{ln}\frac{w+\sqrt{w^{2}-m_{K}^{2}}}{m_{K}})\nonumber\\
&&-72w^{2}\sqrt{m_{\eta}^{2}-w^{2}}(\text{arccos}\frac{-w}{m_{\eta}})\theta(m_{\eta}^{2}-w^{2})
+72w^{2}\sqrt{w^{2}-m_{\eta}^{2}}(i\pi-\text{ln}\frac{w+\sqrt{w^{2}-m_{\eta}^{2}}}{m_{\eta}})\theta(w^{2}-m_{\eta}^{2})].
\end{eqnarray}

Figure 2(n):
\begin{eqnarray}
 V_{KN}^{(1)}&=&\frac{1}{128\pi^{2}f_{K}^{4}}[-40w^{3}+(40w^{3}-15wm_{K}^{2})\text{ln}\frac{m_{K}}{\lambda}
+(16w^{3}-6wm_{\pi}^{2})\text{ln}\frac{m_{\pi}}{\lambda}+(24w^{3}-9wm_{\eta}^{2})\text{ln}\frac{m_{\eta}}{\lambda}\nonumber\\
&&+40w^{2}\sqrt{w^{2}-m_{K}^{2}}\text{ln}\frac{w+\sqrt{w^{2}-m_{K}^{2}}}{m_{K}}
+16w^{2}\sqrt{w^{2}-m_{\pi}^{2}}\text{ln}\frac{w+\sqrt{w^{2}-m_{\pi}^{2}}}{m_{\pi}}\nonumber\\
&&-24w^{2}\sqrt{m_{\eta}^{2}-w^{2}}(\text{arccos}\frac{w}{m_{\eta}})\theta(m_{\eta}^{2}-w^{2})
+24w^{2}\sqrt{w^{2}-m_{\eta}^{2}}(\text{ln}\frac{w+\sqrt{w^{2}-m_{\eta}^{2}}}{m_{\eta}})\theta(w^{2}-m_{\eta}^{2})],
\end{eqnarray}
\begin{eqnarray}
V_{KN}^{(0)}&=&\frac{3}{128\pi^{2}f_{K}^{4}}[(8w^{3}-3wm_{\pi}^{2})\text{ln}\frac{m_{\pi}}{\lambda}-(8w^{3}-3wm_{K}^{2})\text{ln}\frac{m_{K}}{\lambda}+8w^{2}\sqrt{w^{2}-m_{\pi}^{2}}\text{ln}\frac{w+\sqrt{w^{2}-m_{\pi}^{2}}}{m_{\pi}}\nonumber\\
&&-8w^{2}\sqrt{w^{2}-m_{K}^{2}}\text{ln}\frac{w+\sqrt{w^{2}-m_{K}^{2}}}{m_{K}}],
\end{eqnarray}
\begin{eqnarray}
V_{\overline{K}N}^{(1)}=\frac{1}{64\pi^{2}f_{K}^{4}}[-4w^{3}+(8w^{3}-3wm_{K}^{2})\text{ln}\frac{m_{K}}{\lambda}+8w^{2}\sqrt{w^{2}-m_{K}^{2}}\text{ln}\frac{w+\sqrt{w^{2}-m_{K}^{2}}}{m_{K}}],
\end{eqnarray}
\begin{eqnarray}
V_{\overline{K}N}^{(0)}=\frac{3}{64\pi^{2}f_{K}^{4}}[-4w^{3}+(8w^{3}-3wm_{K}^{2})\text{ln}\frac{m_{K}}{\lambda}+8w^{2}\sqrt{w^{2}-m_{K}^{2}}\text{ln}\frac{w+\sqrt{w^{2}-m_{K}^{2}}}{m_{K}}].
\end{eqnarray}

Figures 2(o) and 2(p):
\begin{eqnarray}
 V_{KN}^{(1)}&=&-\frac{1}{288\pi f_{K}^{4}}[3(D+F)(D+3F)m_{\pi}^{3}+(7D^{2}
-6DF+15F^{2})m_{K}^{3}-6F(D-3F)m_{\eta}^{3}],
\end{eqnarray}
\begin{eqnarray}
V_{KN}^{(0)}=-\frac{1}{96\pi f_{K}^{4}}[3(D^{2}-F^{2})m_{\pi}^{3}+3(D-F)^{2}m_{K}^{3}+2D(D-3F)m_{\eta}^{3}],
\end{eqnarray}
\begin{eqnarray}
V_{\overline{K}N}^{(1)}=-\frac{1}{288\pi f_{K}^{4}}[6D(D+F)m_{\pi}^{3}+4(2D^{2}-3DF+3F^{2})m_{K}^{3}+3(D-3F)(D-F)m_{\eta}^{3}],
\end{eqnarray}
\begin{eqnarray}
V_{\overline{K}N}^{(0)}=-\frac{1}{96\pi f_{K}^{4}}[6F(D+F)m_{\pi}^{3}+2(D^{2}+3F^{2})m_{K}^{3}+(9F^{2}-D^{2})m_{\eta}^{3}].
\end{eqnarray}

Figure 2(q):
\begin{eqnarray}
V_{KN}^{(1)}=-\frac{w}{96\pi^{2}f_{K}^{4}}[6(D+F)^{2}m_{\pi}^{2}(1+3\text{ln}\frac{m_{\pi}}{\lambda})
+3(D-F)^{2}m_{K}^{2}(1+3\text{ln}\frac{m_{K}}{\lambda})
+(D-3F)^{2}m_{\eta}^{2}(1+3\text{ln}\frac{m_{\eta}}{\lambda})],
\end{eqnarray}
\begin{eqnarray}
V_{KN}^{(0)}=-\frac{3w}{32\pi^{2}f_{K}^{4}}[(D+F)^{2}m_{\pi}^{2}(1+3\text{ln}\frac{m_{\pi}}{\lambda})
-(D-F)^{2}m_{K}^{2}(1+3\text{ln}\frac{m_{K}}{\lambda})],
\end{eqnarray}
\begin{eqnarray}
V_{\overline{K}N}^{(1)}=\frac{w}{192\pi^{2}f_{K}^{4}}[15(D+F)^{2}m_{\pi}^{2}(1+3\text{ln}\frac{m_{\pi}}{\lambda})
-6(D-F)^{2}m_{K}^{2}(1+3\text{ln}\frac{m_{K}}{\lambda})+(D-3F)^{2}m_{\eta}^{2}(1+3\text{ln}\frac{m_{\eta}}{\lambda})],
\end{eqnarray}
\begin{eqnarray}
V_{\overline{K}N}^{(0)}=\frac{w}{64\pi^{2}f_{K}^{4}}[3(D+F)^{2}m_{\pi}^{2}(1+3\text{ln}\frac{m_{\pi}}{\lambda})
+6(D-F)^{2}m_{K}^{2}(1+3\text{ln}\frac{m_{K}}{\lambda})+(D-3F)^{2}m_{\eta}^{2}(1+3\text{ln}\frac{m_{\eta}}{\lambda})].
\end{eqnarray}

Figure 2(r):
\begin{eqnarray}
V_{KN}^{(1)}&=&\frac{w}{128\pi^{2}f_{K}^{4}}(2m_{\pi}^{2}\text{ln}\frac{m_{\pi}}{\lambda}+7m_{K}^{2}\text{ln}\frac{m_{K}}{\lambda}
+m_{\eta}^{2}\text{ln}\frac{m_{\eta}}{\lambda}),
\end{eqnarray}
\begin{eqnarray}
V_{KN}^{(0)}&=&-\frac{3w}{128\pi^{2}f_{K}^{4}}(m_{\pi}^{2}\text{ln}\frac{m_{\pi}}{\lambda}-m_{K}^{2}\text{ln}\frac{m_{K}}{\lambda}),
\end{eqnarray}
\begin{eqnarray}
V_{\overline{K}N}^{(1)}&=&\frac{w}{256\pi^{2}f_{K}^{4}}(m_{\pi}^{2}\text{ln}\frac{m_{\pi}}{\lambda}-10m_{K}^{2}\text{ln}\frac{m_{K}}{\lambda}-m_{\eta}^{2}\text{ln}\frac{m_{\eta}}{\lambda}),
\end{eqnarray}
\begin{eqnarray}
V_{\overline{K}N}^{(0)}&=&-\frac{3w}{256\pi^{2}f_{K}^{4}}(3m_{\pi}^{2}\text{ln}\frac{m_{\pi}}{\lambda}+6m_{K}^{2}\text{ln}\frac{m_{K}}{\lambda}+m_{\eta}^{2}\text{ln}\frac{m_{\eta}}{\lambda}).
\end{eqnarray}

Figure 2(s):
\begin{eqnarray}
V_{KN}^{(1)}&=&\frac{w}{384\pi^{2}f_{K}^{4}}\{8m_{\pi}^{2}+40m_{K}^{2}-10Q^{2}-2(6m_{\pi}^{2}-Q^{2})\text{ln}\frac{m_{\pi}}{\lambda}-10(6m_{K}^{2}-Q^{2})\text{ln}\frac{m_{K}}{\lambda}\nonumber\\
&&-(4m_{\pi}^{2}-Q^{2})r(m_\pi)\text{ln}|\frac{1+r(m_\pi)}{1-r(m_\pi)}|
-5(4m_{K}^{2}-Q^{2})r(m_K)\text{ln}|\frac{1+r(m_K)}{1-r(m_K)}|\},
\end{eqnarray}
\begin{eqnarray}
V_{KN}^{(0)}&=&\frac{w}{384\pi^{2}f_{K}^{4}}\{-24m_{\pi}^{2}+24m_{K}^{2}+6(6m_{\pi}^{2}-Q^{2})\text{ln}\frac{m_{\pi}}{\lambda}-6(6m_{K}^{2}-Q^{2})\text{ln}\frac{m_{K}}{\lambda}\nonumber\\
&&+3(4m_{\pi}^{2}-Q^{2})r(m_\pi)\text{ln}|\frac{1+r(m_\pi)}{1-r(m_\pi)}|
-3(4m_{K}^{2}-Q^{2})r(m_K)\text{ln}|\frac{1+r(m_K)}{1-r(m_K)}|\},
\end{eqnarray}
\begin{eqnarray}
V_{\overline{K}N}^{(1)}&=&\frac{w}{384\pi^{2}f_{K}^{4}}\{8m_{\pi}^{2}-32m_{K}^{2}+5Q^{2}-2(6m_{\pi}^{2}-Q^{2})\text{ln}\frac{m_{\pi}}{\lambda}+8(6m_{K}^{2}-Q^{2})\text{ln}\frac{m_{K}}{\lambda}\nonumber\\
&&-(4m_{\pi}^{2}-Q^{2})r(m_\pi)\text{ln}|\frac{1+r(m_\pi)}{1-r(m_\pi)}|
+4(4m_{K}^{2}-Q^{2})r(m_K)\text{ln}|\frac{1+r(m_K)}{1-r(m_K)}|\},
\end{eqnarray}
\begin{eqnarray}
V_{\overline{K}N}^{(0)}&=&\frac{w}{384\pi^{2}f_{K}^{4}}\{-24m_{\pi}^{2}-48m_{K}^{2}+15Q^{2}+6(6m_{\pi}^{2}-Q^{2})\text{ln}\frac{m_{\pi}}{\lambda}+12(6m_{K}^{2}-Q^{2})\text{ln}\frac{m_{K}}{\lambda}\nonumber\\
&&+3(4m_{\pi}^{2}-Q^{2})r(m_\pi)\text{ln}|\frac{1+r(m_\pi)}{1-r(m_\pi)}|
+6(4m_{K}^{2}-Q^{2})r(m_K)\text{ln}|\frac{1+r(m_K)}{1-r(m_K)}|\}.
\end{eqnarray}

Figure 2(t):
\begin{eqnarray}
V_{KN}^{(1)}&=&\frac{1}{576\pi f_{K}^{4}}\{3(D+F)^{2}[3 m_{\pi}^{3}-3m_{\pi}Q^{2}+36\pi m_{\pi}^{2}Q^{2}K_{0}(m_{\pi})+\frac{2w}{\pi}H_{0}(m_{\pi})]+2(7D^{2}-6DF+15F^{2})\nonumber\\
&&\times[m_{K}^{3}-m_{K}Q^{2}+12\pi m_{K}^{2}Q^{2}K_{0}(m_{K})+\frac{w}{\pi}H_{0}(m_{K})]+(D-3F)^{2}[3m_{\eta}^{3}+4m_{\eta}m_{K}^{2}-3m_{\eta}Q^{2}\nonumber\\
&&-4\pi(8m_{K}^{2}-9Q^{2})m_{\eta}^{2}K_{0}(m_{\eta})]+(D+F)(3F-D)[\frac{1}{2}(m_{\pi}+m_{\eta})(15m_{\pi}^{2}+21m_{\eta}^{2}-18m_{\pi}m_{\eta}-9Q^{2})\nonumber\\
&&-2\pi(12m_{\eta}^{4}-60m_{\pi}^{2}m_{\eta}^{2}+30m_{\pi}^{2}Q^{2}
-6m_{\eta}^{2}Q^{2}+3Q^{4})H(m_{\pi},m_{\eta})-2\pi(12m_{\pi}^{4}-
36m_{\pi}^{2}m_{\eta}^{2}+30m_{\eta}^{2}Q^{2}\nonumber\\
&&-6m_{\pi}^{2}Q^{2}+3Q^{4}-24m_{\eta}^{4})H(m_{\eta},m_{\pi})]\},
\end{eqnarray}
\begin{eqnarray}
V_{KN}^{(0)}&=&\frac{1}{576\pi f_{K}^{4}}\{3(D+F)^{2}[3 m_{\pi}^{3}-3m_{\pi}Q^{2}+36\pi m_{\pi}^{2}Q^{2}K_{0}(m_{\pi})-\frac{6w}{\pi}H_{0}(m_{\pi})]+18(D-F)^{2}\nonumber\\
&&\times[m_{K}^{3}-m_{K}Q^{2}+12\pi m_{K}^{2}Q^{2}K_{0}(m_{K})+\frac{w}{\pi}H_{0}(m_{K})]+(D-3F)^{2}[3m_{\eta}^{3}+4m_{\eta}m_{K}^{2}-3m_{\eta}Q^{2}\nonumber\\
&&-4\pi(8m_{K}^{2}-9Q^{2})m_{\eta}^{2}K_{0}(m_{\eta})]+3(D+F)(D-3F)[\frac{1}{2}(m_{\pi}+m_{\eta})(15m_{\pi}^{2}+21m_{\eta}^{2}-18m_{\pi}m_{\eta}-9Q^{2})\nonumber\\
&&-2\pi(12m_{\eta}^{4}-60m_{\pi}^{2}m_{\eta}^{2}+30m_{\pi}^{2}Q^{2}
-6m_{\eta}^{2}Q^{2}+3Q^{4})H(m_{\pi},m_{\eta})-2\pi(12m_{\pi}^{4}
-36m_{\pi}^{2}m_{\eta}^{2}+30m_{\eta}^{2}Q^{2}\nonumber\\
&&-6m_{\pi}^{2}Q^{2}+3Q^{4}-24m_{\eta}^{4})H(m_{\eta},m_{\pi})]\},
\end{eqnarray}
\begin{eqnarray}
V_{\overline{K}N}^{(1)}&=&\frac{1}{576\pi f_{K}^{4}}\{3(D+F)^{2}[3 m_{\pi}^{3}-3m_{\pi}Q^{2}+36\pi m_{\pi}^{2}Q^{2}K_{0}(m_{\pi})+\frac{2w}{\pi}H_{0}(m_{\pi})]+8(2D^{2}-3DF+3F^{2})\nonumber\\
&&\times[m_{K}^{3}-m_{K}Q^{2}+12\pi m_{K}^{2}Q^{2}K_{0}(m_{K})-\frac{w}{\pi}H_{0}(m_{K})]+(D-3F)^{2}[3m_{\eta}^{3}+4m_{\eta}m_{K}^{2}-3m_{\eta}Q^{2}\nonumber\\
&&-4\pi(8m_{K}^{2}-9Q^{2})m_{\eta}^{2}K_{0}(m_{\eta})]+(D+F)(D-3F)[\frac{1}{2}(m_{\pi}+m_{\eta})(15m_{\pi}^{2}+21m_{\eta}^{2}-18m_{\pi}m_{\eta}-9Q^{2})\nonumber\\
&&-2\pi(12m_{\eta}^{4}-60m_{\pi}^{2}m_{\eta}^{2}+30m_{\pi}^{2}Q^{2}
-6m_{\eta}^{2}Q^{2}+3Q^{4})H(m_{\pi},m_{\eta})-2\pi(12m_{\pi}^{4}
-36m_{\pi}^{2}m_{\eta}^{2}+30m_{\eta}^{2}Q^{2}\nonumber\\
&&-6m_{\pi}^{2}Q^{2}+3Q^{4}-24m_{\eta}^{4})H(m_{\eta},m_{\pi})]\},
\end{eqnarray}
\begin{eqnarray}
V_{\overline{K}N}^{(0)}&=&\frac{1}{576\pi f_{K}^{4}}\{3(D+F)^{2}[3 m_{\pi}^{3}-3m_{\pi}Q^{2}+36\pi m_{\pi}^{2}Q^{2}K_{0}(m_{\pi})-\frac{6w}{\pi}H_{0}(m_{\pi})]+12(D^{2}+3F^{2})\nonumber\\
&&\times[m_{K}^{3}-m_{K}Q^{2}+12\pi m_{K}^{2}Q^{2}K_{0}(m_{K})-\frac{w}{\pi}H_{0}(m_{K})]+(D-3F)^{2}[3m_{\eta}^{3}+4m_{\eta}m_{K}^{2}-3m_{\eta}Q^{2}\nonumber\\
&&-4\pi(8m_{K}^{2}-9Q^{2})m_{\eta}^{2}K_{0}(m_{\eta})]+3(D+F)(3F-D)[\frac{1}{2}(m_{\pi}+m_{\eta})(15m_{\pi}^{2}+21m_{\eta}^{2}-18m_{\pi}m_{\eta}-9Q^{2})\nonumber\\
&&-2\pi(12m_{\eta}^{4}-60m_{\pi}^{2}m_{\eta}^{2}+30m_{\pi}^{2}Q^{2}
-6m_{\eta}^{2}Q^{2}+3Q^{4})H(m_{\pi},m_{\eta})-2\pi(12m_{\pi}^{4}
-36m_{\pi}^{2}m_{\eta}^{2}+30m_{\eta}^{2}Q^{2}\nonumber\\
&&-6m_{\pi}^{2}Q^{2}+3Q^{4}-24m_{\eta}^{4})H(m_{\eta},m_{\pi})]\},
\end{eqnarray}

where
\begin{eqnarray}
K_{0}(m)=-\frac{1}{8\pi\sqrt{-Q^2}}\text{arctan}\frac{\sqrt{-Q^2}}{2m},
\end{eqnarray}
\begin{eqnarray}
H_{0}(m)=-m^{2}+\frac{Q^{2}}{6}-5m^{2}\text{ln}\frac{m}{\lambda}+(2m^{2}+\frac{Q^{2}}{4})[1-2\text{ln}\frac{m}{\lambda}-r(m)\text{ln}|\frac{1+r(m)}{1-r(m)}|],
\end{eqnarray}
\begin{eqnarray}
H(m_{1},m_{2})=\frac{1}{32\pi^{2}\sqrt{-Q^{2}}}[F_{1}(y_{1})+F_{1}(y_{2})+F_{2}(x_{1})+F_{2}(x_{2})-F_{3}(z_{1})-F_{3}(z_{2})],
\end{eqnarray}
with
\begin{eqnarray}
 &F_{1}(x)=\text{Li}_{2}(\frac{z_{0}+2Q^{2}}{z_{0}-2\sqrt{-Q^{2}}x})-\text{Li}_{2}(\frac{z_{0}}{z_{0}-2\sqrt{-Q^{2}}x}),\nonumber\\
&F_{2}(x)=-\text{Li}_{2}(\frac{z_{0}+2Q^{2}}{z_{0}-2\sqrt{-Q^{2}}x})-\frac{1}{2}\text{ln}^{2}(\frac{z_{0}-2\sqrt{-Q^{2}}x}{\lambda^{2}}),\nonumber\\
 &F_{3}(x)=-\text{Li}_{2}(\frac{z_{0}}{z_{0}-2\sqrt{-Q^{2}}x})-\frac{1}{2}\text{ln}^{2}(\frac{z_{0}-2\sqrt{-Q^{2}}x}{\lambda^{2}}),\nonumber\\
&z_{0}=m_{2}^{2}-m_{1}^{2}-Q^{2},\quad\quad x_{1,2}=\sqrt{-Q^{2}}\pm im_{2},\quad\quad z_{1,2}=\pm im_{1},\nonumber\\
&y_{1,2}=-\frac{1}{2\sqrt{-Q^{2}}}(m_{1}^{2}-m_{2}^{2}+Q^{2}
\pm\sqrt{[Q^{2}-(m_{1}+m_{2})^{2}][Q^{2}-(m_{1}-m_{2})^{2}]}).
\end{eqnarray}
The dilogarithm or Spence function (polylogarithm function) is defined by
\begin{eqnarray}
\text{Li}_{2}(x)=-\int_{0}^{1}dt\frac{\text{ln}(1-xt)}{t}.
\end{eqnarray}
External leg (wave function) renormalization (including the contributions resulting from renormalizing $f$ to $f_{K}$ in the leading-order terms)
\begin{eqnarray}
V_{KN}^{(1)}=\frac{1}{3f_{K}^{2}}[-3w+(D^{2}+3F^{2})\frac{zq^{2}}{w}](\delta Z_{N}-2\delta Z_{K}),
\end{eqnarray}
\begin{eqnarray}
W_{KN}^{(1)}=-\frac{1}{3wf_{K}^{2}}(D^{2}+3F^{2})(\delta Z_{N}-2\delta Z_{K}),
\end{eqnarray}
\begin{eqnarray}
\label{eq13}
V_{KN}^{(0)}(q)=\frac{(2D^{2}-6DF)q^{2}z}{3wf_{K}^{2}}(\delta Z_{N}-2\delta Z_{K}),
\end{eqnarray}
\begin{eqnarray}
\label{eq14}
W_{KN}^{(0)}(q)=-\frac{2D^{2}-6DF}{3wf_{K}^{2}}(\delta Z_{N}-2\delta Z_{K}),
\end{eqnarray}
\begin{eqnarray}
\label{eq15}
V_{\overline{K}N}^{(1)}(q)=\frac{1}{2f_{K}^{2}}[w-(D-F)^{2}\frac{q^{2}z}{w}](\delta Z_{N}-2\delta Z_{K}),
\end{eqnarray}
\begin{eqnarray}
\label{eq16}
W_{\overline{K}N}^{(1)}(q)=-\frac{(D-F)^{2}}{2wf_{K}^{2}}(\delta Z_{N}-2\delta Z_{K}),
\end{eqnarray}
\begin{eqnarray}
\label{eq17}
V_{\overline{K}N}^{(0)}(q)=\frac{1}{6f_{K}^{2}}[9w-(D+3F)^{2}\frac{q^{2}z}{w}](\delta Z_{N}-2\delta Z_{K}),
\end{eqnarray}
\begin{eqnarray}
\label{eq18}
W_{\overline{K}N}^{(0)}(q)=-\frac{(D+3F)^{2}}{6wf_{K}^{2}}(\delta Z_{N}-2\delta Z_{K}),
\end{eqnarray}
where
\begin{eqnarray}
\delta Z_{N}&=&-\frac{1}{96\pi^{2}f_{K}^{2}}[9(D+F)^{2}m_{\pi}^{2}(1+3\text{ln}\frac{m_{\pi}}{\lambda})+(10D^{2}+18F^{2}-12DF)m_{K}^{2}(1+3\text{ln}\frac{m_{K}}{\lambda})\nonumber\\
&&+(D-3F)^{2}m_{\eta}^{2}(1+3\text{ln}\frac{m_{\eta}}{\lambda})],\nonumber\\
\delta Z_{K}&=&\frac{1}{32\pi^{2}f_{K}^{2}}(m_{\pi}^{2}\text{ln}\frac{m_{\pi}}{\lambda}+2m_{K}^{2}\text{ln}\frac{m_{K}}{\lambda}+m_{\eta}^{2}\text{ln}\frac{m_{\eta}}{\lambda}).
\end{eqnarray}
\end{widetext}

\section{Threshold T-matrices}
\label{Threshold T-matrices}
We take the relations between the threshold T-matrices and the scattering lengths: $T_{KN,th}^{(0,1)}=4\pi(1+m_{K}/M_{N})a_{KN}^{(0,1)}$ and $T_{\overline{K}N,th}^{(0,1)}=4\pi(1+m_{K}/M_{N})a_{\overline{K}N}^{(0,1)}$. The threshold T-matrix can be obtained through $\vec{q}=0$ when the corresponding diagrams are calculated. Here, the one-loop order threshold T- matrices indicated by $\mathcal{O}(q^{3})$ are the same as in Ref.~\cite{liu20071} and not given obviously. The resulting threshold T-matrices are given by
\onecolumngrid
\begin{eqnarray}
\label{B1}
T_{KN,th}^{(1)}=-\frac{m_{K}}{f_{K}^{2}}[1+\frac{m_{K}}{6M_{0}}(D^{2}+3F^{2})+m_{K}(4b_{D}+4b_{0}
+C_{1}+C_{2})]+\mathcal{O}(q^{3}),
\end{eqnarray}
\begin{eqnarray}
\label{B2}
T_{KN,th}^{(0)}=\frac{m_{K}^{2}}{f_{K}^{2}}[\frac{D}{3M_{0}}(3F-D)+(4b_{F}-4b_{0}
+C_{4}+C_{5})]+\mathcal{O}(q^{3}),
\end{eqnarray}
\begin{eqnarray}
\label{B3}
T_{\overline{K}N,th}^{(1)}=\frac{m_{K}}{f_{K}^{2}}[\frac{1}{2}-\frac{m_{K}}{4M_{0}}(D-F)^{2}+m_{K}(2b_{F}-2b_{D}-4b_{0})
-\frac{m_{K}}{2}(C_{1}+C_{2}-C_{4}-C_{5})]+\mathcal{O}(q^{3}),
\end{eqnarray}
\begin{eqnarray}
\label{B4}
T_{\overline{K}N,th}^{(0)}=\frac{m_{K}}{f_{K}^{2}}[\frac{3}{2}-\frac{m_{K}}{12M_{0}}(D+3F)^{2}-2m_{K}(3b_{D}+b_{F}+2b_{0})
-\frac{m_{K}}{2}(3C_{1}+3C_{2}+C_{4}+C_{5})]+\mathcal{O}(q^{3}).
\end{eqnarray}
\nocite{*}

\twocolumngrid
\providecommand{\noopsort}[1]{}\providecommand{\singleletter}[1]{#1}%
\bibliography{apssamp}

\end{document}